\documentclass[aps,prb,twocolumn,showpacs,floatfix,a4paper]{revtex4}
\usepackage{graphicx}
\usepackage[LGR,T1]{fontenc}
\usepackage{geometry}
\usepackage{units}
\usepackage{amsmath}
\usepackage{amssymb}
\usepackage{color}

\usepackage{natbib}
\usepackage{esint}
\usepackage{bm}
\usepackage{ulem}
\usepackage[colorlinks=true, citecolor=blue]{hyperref}
\setcitestyle{journalcolor= blue}

\begin{document}

	\title{Quantum Phases of a Weakly Disordered Josephson Ladder}

	\author{Eyal Walach}
	\author{Efrat Shimshoni}

	\affiliation{Department of Physics, Jack and Pearl Resnick Institute and the Institute of Nanotechnology and Advanced Materials,
		Bar-Ilan University, Ramat-Gan 52900, Israel}

\date{\today}

\begin{abstract}
The interplay of interactions and disorder in low-dimensional superconductors supports the formation of multiple quantum phases, as possible instabilities of the Superconductor-Insulator Transition (SIT) at a singular quantum critical point. We explore a one-dimensional model which exhibits such variety of phases in the strongly quantum fluctuations regime. Specifically, we study the effect of weak disorder on a two-leg Josephson ladder with comparable Josephson and charging energies ($E_J\sim E_C$). An additional key feature of our model is the requirement of perfect $\mathbb{Z}_2$-symmetry, respected by all parameters including the disorder. Using a perturbative renormalization-group (RG) analysis, we derive the phase diagram and identify at least one intermediate phase between a full-fledged superconductor and a disorder-dominated insulator. Most prominently, for repulsive interactions on the rungs we identify two distinct mixed phases: in both of them the longitudinal charge mode is a gapless superconductor, however one phase exhibits a dipolar charge density order on the rungs, while the other is disordered. This latter phase is characterized by coexisting superconducting (phase-locked) and charge-ordered rungs, and encompasses the potential of evolving into a Grifith's phase characteristic of the random-field Ising model in the strong disorder limit.

\end{abstract}

	\maketitle

\section{Introduction and Principal Results}
\label{sec:Introduction}

The superconductor-insulator transition (SIT) observed in thin layers or wires of superconducting (SC)
materials is a dramatic manifestation of quantum fluctuations enhanced by the low-dimensionality \cite{1 Hebard 93,1 Sondhi Girvin 97,1 Goldman Markovic 98}. Its most prominent signature is a drastic change in the electric resistance at low temperatures $T\to0$, which switches from zero to infinity upon tuning of
a non-thermal parameter (e.g. a magnetic field, a reduced layer thickness, gating etc.) beyond a critical value. This phenomenon exemplifies a quantum phase transition (QPT) \cite{3 Sachdev 19}: a fundamental change in the nature of the ground state across a $T=0$ critical point.

The onset of a SIT does not necessarily involve the breaking of Cooper pairs: it has been seen in Josephson arrays, granular systems and disordered metals where superconductivity persists locally even in the insulating phase. In such systems, the underlying mechanism
is rather dominated by the combined effects of repulsive interactions and disorder, which tend to imped
long-range phase-coherence between SC islands in favor of a charge-localized phase. This mechanism is
well-captured by interacting Bosons models, or equivalently Josephson arrays \cite{1 Hebard 93,18 Fisher-Weichman 89,18 Vojta-Crewse 16,PRLB Jul18 32 Fisher 90,PRLB Jul18 37 Fisher-Grinstein 90,PRLB Jul18 38 Sorensen-Wallin 92,PRLB Jul18 39 Cha-Girvin 39,PRLB Jul18 40 Prokof'ev-Svistunov,Altman Kafri Polkovnikov Refael 10}. In the latter, the competition between a repulsive interaction
and the superconducting stiffness is tunable by the ratio of two energy scales -- the charging
energy $E_{c}$ and Josephson energy $E_{J}$. The SIT occurs at a critical value where $E_c/E_J\sim 1$, corresponding to maximal phase-charge uncertainty.

Disorder is an additional ingredient, associated
with the presence of random charge impurities and/or spatial fluctuations in
$E_{c}/E_{J}$. Its interplay with the interactions may introduce a richer set of quantum phases, separated by
more than one critical point. Indeed, extensive
studies have suggested a variety of distinct insulating phases including, e.g., a ``Bose/Mott glass" \cite{18 Fisher-Weichman 89,18 Vojta-Crewse 16}. A more intriguing possibility is the emergence of
an intermediate metallic phase \cite{19 Phillips 03,19 Kapitulnik-Kivelson arXiv,20 Mulligan-Raghu 16,20 Goldman-Mulligan 17} near the putative SIT critical point. Alternatively, a mixed phase with coexisting SC and charge density correlations may form in this strongly fluctuating regime.

\begin{figure*}
	\includegraphics[width=18cm,height=!]{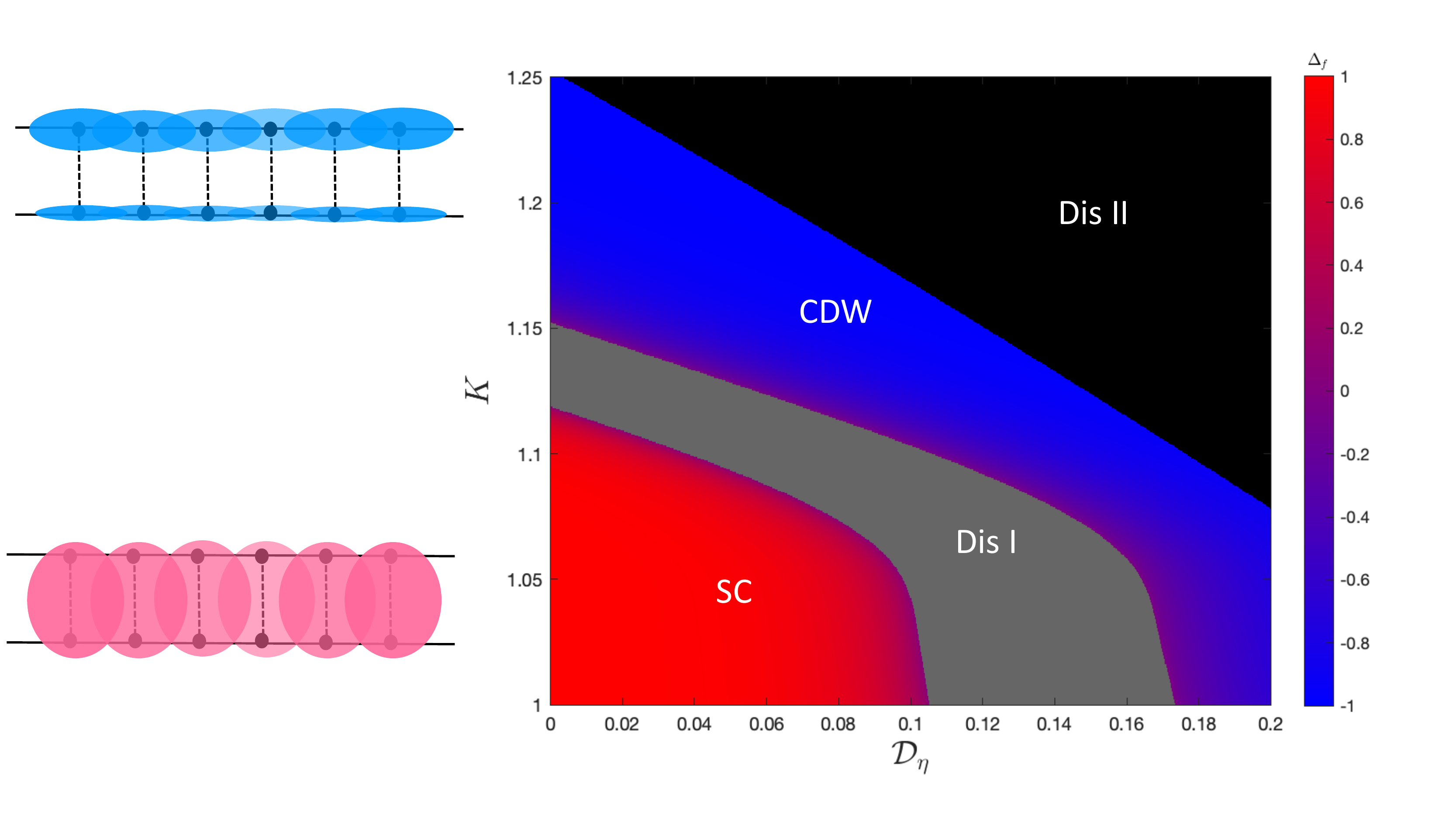}

	\caption{Left panel: Illustration of the charge configuration on the ladder in the ordered phases; the top panel depicts the CDW-ordered phase, and the bottom panel the SC phase. Right panel: Phase diagram as a function of the Luttinger parameter $K=\sqrt{K_{_+} K_{_-}}$ and disorder $\mathcal{D}_\eta$, for $V>0$ (see text); here $\nicefrac{K_{_+}}{K_{_-}}=1.02$, $u_{_+}=u_{_-}=1$, $\mathcal{D}_{g,-}=0.003$, $g_\theta=0.3$, $g_\phi=0.03$, $\mathcal{D}_{U,s}=0$; the color scale denotes $\Delta(\ell_f)$ obtained from the RG flow up to $\ell_f=20$. The grey region (Dis I) corresponds to the intermediate disordered phase where $\mathcal{D}_-$ is relevant, but the symmetric mode maintains a LL behavior; the black region (Dis II) corresponds to the disordered insulator where $\mathcal{D}_g$ is relevant.}
\label{fig:phase_diagram}
\end{figure*}

In the present paper, we show that several mixed phases are supported in a relatively simple model for a strongly fluctuating SC device.
We consider a weakly-disordered two-legged Josephson ladder, focusing on the quantum fluctuations regime where both intra and inter-leg charge interactions are comparable to the Josephson coupling on the same links ($E_c\sim E_J$). A crucial property of our model is a perfect $\mathbb{Z}_2$-symmetry, respected by all parameters including the disorder; notably, this is easier to achieve when the leg index represents a discrete degree of freedom other than real-space separation. Contrary to earlier studies of Bosonic ladders \cite{22 Orignac-Giamarchi 98,22 Dhar-Maji 12,22 Tokuna-Georges 14,22 Ristivojevic-Petkovic 14}, a natural description of the system under these conditions involves a weak coupling between Fermionic and Bosonic sectors. Utilizing a perturbative renormalization-group (RG) analysis to explore the $T=0$ phase diagram, we find evidence for a variety of intermediate phases between a full-fledged superconductor and a disorder-dominated insulator (see Fig. \ref{fig:phase_diagram}). Most prominently, we identify two distinct mixed phase where SC correlations coexist with charge-ordering: one exhibits long-range CDW order, and one (marked Dis I in the figure) is disordered. We conjecture that in the strong disorder limit, this latter phase evolves into a Grifith's phase characteristic of the random-field Ising model \cite{randomIsing}.

A key feature of the Bosonic ladder in the clean limit is the separability of the low-energy degrees of freedom into two independent sectors: the symmetric (longitudinal) and antisymmetric (transverse) modes. Each of the sectors may undergo a QPT of distinct type, associated with the breaking of $U(1)$ and $\mathbb{Z}_2$ symmetries, respectively. In the quantum fluctuations regime, the antisymmetric mode (best represented in terms of nearly-free Fermions) exhibits an Ising-type QPT \cite{28 Atzmon Shimshoni 11,Mazo2014}. Under the extra assumption of commensurate Boson density on the lattice, the symmetric mode exhibits a Berezinskii-Kosterlitz-Thouless (BKT) transition \cite{KosterlitzThouless,Berezinskii} from a Luttinger Liquid (LL) to a Mott insulator. Consequently, as a common parameter is tuned (e.g. $E_{c}/E_{J}$ on the legs), an intermediate phase can emerge between the two separate critical points, with coexisting charge density wave (CDW) order of the global charge and a SC order with inter-leg phase locking.

What is the fate of this intermediate phase in the presence of disorder? One possible scenario is the shrinking of this phase and recovery of a singular SIT critical point. On the other hand, disorder supports the formation of inhomogeneity and thus stabilize the coexistence of spatially separate regions with distinct local order parameters \cite{Nandini_etal}. Specifically in our case, two different types of disorder effects should be considered. The first type, resulting mainly from inhomogeneities in the charging and Josephson energies, maintains the Fermionic and Bosonic sectors independent, but possibly alters the nature and position of the respective critical points. The other type of disorder may introduce coupling terms between the two sectors which may profoundly change the critical behavior of the combined Boson-Fermion system: e.g., turning the continuous Ising transition into first order, or generate a novel multi-critical point \cite{Sitte,29 Huijse Bauer 15,29 Alberton Ruhman 17}. A dominant effect of the latter type stems from the presence of random impurities which induce spatial fluctuations in the chemical potential along the ladder. As detailed below, we analyze the interplay between both types of disorder, and derive their combined effect on the phase diagram.

The rest of this paper is organized as follows: in Sec. \ref{sec:model} we introduce the model; in Sec. \ref{sec:RGanalysis} we present the RG analysis and main results; our concluding remarks are summarized in Sec. \ref{sec:discussion}. Finally, Appendices \ref{app:H_0} through \ref{app:OtherDis} are devoted to technical details of our analysis.

\section{The model}
\label{sec:model}

As a starting point, we consider the clean limit of a two-leg Bosonic ladder described by the Hamiltonian
\begin{equation}
H_0=\int dx \left [\mathcal{H}_{1}+\mathcal{H}_{2}+\mathcal{H}_{int}\right]\label{eq:CleanLimitHamStructure}
\end{equation}
where $x$ is a continuous coordinate along the leg direction (in units of the lattice constant $a$), and the local terms are given by
\begin{align}
\mathcal{H}_{\nu} & =\frac{1}{2}\left[U\rho_{\nu}^{2}+\rho_{s}\left(\partial_{x}\phi_{\nu}\right)^{2}\right]\,\,\left(\nu=1,2\right)\label{eq:CleanLimitLegHam}\\
\mathcal{H}_{int} & =\left[-Jcos\left(\phi_{1}-\phi_{2}\right)+V\rho_{1}\rho_{2}\right]\; .\label{eq:CleanLimitIntHam}
\end{align}
Here $U$ is the charge interaction and $\rho_s$ the superfluid stiffness within each leg, $J$ is the inter-leg Josephson coupling and $V$ is the inter-leg charge interaction; $\rho_\nu(x)$, $\phi_\nu(x)$ are respectively the local charge-density-fluctuation and phase operators on leg $\nu$.  We further use the expansion \cite{27 Giamarchi 04}
\begin{equation}
\rho_\nu=-\frac{1}{\pi} \partial_x \theta_\nu + \rho_0 \sum_{p\in\mathbb{Z}\setminus 0} e^{i\cdot 2p \cdot (\theta_\nu-\pi \rho_0 x)}
\label{eq:rho_nu}
\end{equation}
where $\partial_x\theta_\nu$ and $\phi_\nu$ are canonical conjugates, and $\rho_0$ denotes the uniform background charge density.

Since $H_0$ obeys a $\mathbb{Z}_2$ symmetry to exchange between the legs, its low-energy approximation can be decomposed into independent symmetric ($+$) and antisymmetric ($-$) sectors using the transformation
\begin{equation}
\theta_\pm=\frac{\theta_1\pm\theta_2}{\sqrt{2}}\; ,\quad \phi_\pm=\frac{\phi_1\pm\phi_2}{\sqrt{2}}\; .
\label{eq:SectorDefinition}
\end{equation}
Accounting for the leading terms in Eq. (\ref{eq:rho_nu}) [see App. \ref{app:H_0} for details], this yields $H_0=H_+ + H_-$ where each subsystem independently exhibits a QPT tunable by a common parameter of $H_0$, e.g. $K\propto\sqrt{\frac{U}{\rho_s}}$.

The symmetric part $H_+$ is a sine-Gordon (SG) model:
\begin{align}
H_+ & =\int dx \left[\mathcal{H}_{LL}^{(+)} + g \cos\left(\sqrt{8}\theta_+ -2\pi \rho_0 x\right)\right]\; , \label{eq:SymmetricSector}\\
\mathcal{H}_{LL}^{(+)} & =  \frac{u_{_+}}{2\pi} \left(K_{_+}(\partial_x \theta_+)^2 + \frac{1}{K_{_+}} (\partial_x \phi_+)^2\right)  \nonumber
\end{align}
which exhibits a Luttinger liquid (LL) behavior corresponding to a gapless plasmon mode for generic values of $\rho_0$. However close to integer filling of the underlying lattice, a transition to a Mott insulator occurs when the Luttinger parameter $K_{_+}\sim\sqrt{\frac{(U+V)}{\rho_s}}$ exceeds a critical value; this is a SIT where the SC phase exhibits only a quasi long-range order.

In contrast, $H_-$ describing the antisymmetric mode
is a self-dual SG model (SDSG) \cite{SDSG}
\begin{align}
H_- & = \int dx \left[\mathcal{H}_{LL}^{(-)}  -g_\phi\cos\sqrt{2}\phi_- + g_\theta \cos\sqrt{8}\theta_-\right] \; ,  \label{eq:AntiSymmetricSDSG}\\
\mathcal{H}_{LL}^{(-)} &  =   \frac{u_{_-}}{2\pi} \left(K_{_-}(\partial_x \theta_-)^2 + \frac{1}{K_{_-}} (\partial_x \phi_-)^2\right)  \nonumber
\label{eq:AntiSymmetricSector}
\end{align}
in which the competing phase-locking and charge-locking cosine terms arise from the corresponding two terms of Eq. (\ref{eq:CleanLimitIntHam}).
In a wide range of parameters surrounding the self-duality point $K_{_-}=2$, $g_\phi=g_\theta$ (accessible for $U\sim \rho_s$ in Eq. (\ref{eq:CleanLimitLegHam}) and $J\sim V\rho_0^2$), both of them are simultaneously relevant and the SDSG
is effectively described as two independent transverse-field Ising models, one of which is highly massive \cite{27 Gogolin Nersesyan 98,Mazo2014}. The low-energy description is therefore
given in terms of a single pair of Majorana fields $\xi_R$, $\xi_L$:
\begin{equation}
H_-=  \int dx \left(\xi_R(-iu_{_-}\partial_x)\xi_R - \xi_L(-iu_{_-}\partial_x)\xi_L-i\Delta \xi_R \xi_L\right)\label{eq:AntiSymmetricSector}
\end{equation}
which indicates an Ising-type transition when the gap $\Delta$ changes sign. This can be interpreted as a SIT as well: the $\Delta>0$ phase (realized when $J$ in Eq. (\ref{eq:CleanLimitIntHam}) is sufficiently larger than $V\rho_0^2$) is phase-locked (i.e. SC), while $\Delta<0$ (corresponding to the opposite case) is a Mott insulator; both phases are long-range ordered (with a gap $|\Delta|$). Note that the nature of CDW order in the insulator depends on the sign of $V$: for $V>0$, dipoles are formed on the rungs ($\theta_-=\pm\pi/\sqrt{8}$), while $V<0$ favors equal charges on the two legs ($\theta_-=0$).

We now introduce disorder resulting from random $x$-dependent variations in the various parameters of the model. We distinguish two types of disorder, as detailed below.

(a) {\it Particle-hole preserving disorder.} We first consider randomness arising from spatial inhomogeneities in the parameter $U$, $\rho_s$, $J$ and $V$ of the original model Eqs. (\ref{eq:CleanLimitLegHam}), (\ref{eq:CleanLimitIntHam}), related to the charging and Josephson energies on the legs and rungs of the ladder. Such corrections to the Hamiltonian do not couple linearly to the density operators, and hence do not violate particle-hole symmetry when the chemical potential adjusts $\rho_0$ to a commensurate filling. At the same time this type of disorder maintains the $\mathbb{Z}_2$-symmetry of the model; hence it does not couple the symmetric and antisymmetric sectors and is ultimately manifested as randomness in the parameters of $H_+$, $H_-$ [Eqs. (\ref{eq:SymmetricSector}), (\ref{eq:AntiSymmetricSector})]. Notably, since the disorder is space-dependent but not time-dependent, it breaks the Lorentz symmetry characterizing both low-energy degrees of freedom; hence (as we show explicitly in the next Section) all the parameters including the velocities $u_{_\pm}$ flow under RG.

The disorder in the symmetric sector is introduced as $x$-dependent corrections to the parameters $K_{_+}$, $u_{_+}$ and $g$; all of these can be assumed to originate from a term $\delta\mathcal{H}_{\nu}(x)$ of the form Eq. (\ref{eq:CleanLimitLegHam}) with random charging energy $\delta U(x)$ and superfluid stiffness $\delta \rho_s(x)$. We further assume that these random corrections are the same on both legs $\nu=1,2$ and correspond to a static "white noise" characterized by the disorder averages
\begin{align}
\left<\left<\delta U(x)\right>\right> =0 \; ,\quad & \left<\left<\delta \rho_s(x)\right>\right>=0 \nonumber \\
\left<\left<\delta U(x)\delta U(x^\prime)\right>\right> &= D_{_U} \delta(x-x^\prime) \label{eq:D_Us_def} \\
\left<\left<\delta \rho_s(x)\delta \rho_s(x^\prime)\right>\right> &= D_{s} \delta(x-x^\prime)\; . \nonumber
\label{eq:QuadraticDisorder}
\end{align}
As shown in the next Section, these disorder terms renormalize the parameters $K_{_+}$, $u_{_+}$ of the quadratic part $\mathcal{H}_{LL}^{(+)}$ in Eq. (\ref{eq:SymmetricSector}), but are irrelevant under RG. More significant is their effect on the cosine term, which we maintain as an independent disorder term associated with random corrections to $g$:
\begin{equation}
H_g  =\frac{1}{2}\int dx \left(\delta g(x) e^{i\sqrt{8} \theta_+}+\delta g^\ast(x) e^{-i\sqrt{8} \theta_+}\right)
\label{eq:Hdis_g}
\end{equation}
where the complex parameter $\delta g(x)$ contains the oscillatory phase shift of $\theta_+$, and is characterized by the disorder averages
\begin{align}
\left<\left<\delta g(x)\delta g(x^\prime)\right>\right> &=\left<\left<\delta g(x)\right>\right>=0 \; , \nonumber \\
\left<\left<\delta g(x)\delta g^\ast(x^\prime)\right>\right> &= D_{g} \delta(x-x^\prime)\; .
\label{eq:D_g_def}
\end{align}

In the anti-symmetric sector, the disorder characterized by $D_{_U}$, $D_s$ [Eq. (\ref{eq:D_Us_def})] combined with random fluctuations in the rung-interactions $J$, $V$ generate $x$-dependence in all the parameters of Eq. (\ref{eq:AntiSymmetricSDSG}). However, within the regime of parameters where the low-energy theory for $H_-$ is captured by Eq. (\ref{eq:AntiSymmetricSector}), we encode their most prominent contribution in a single additional disorder parameter corresponding to spatially-dependent corrections to the mass $\Delta$:
\begin{align}
\left<\left<\delta \Delta(x)\right>\right>=0 \; ,& \nonumber \\
\left<\left<\delta \Delta(x)\delta \Delta(x^\prime)\right>\right> &= D_{_-} \delta(x-x^\prime)\; .
\label{eq:D_-_def}
\end{align}
Mapping to the Ising model, $\delta \Delta(x)$ can be interpreted as a random transverse field.

(b) {\it Disordered chemical potential.} We next consider randomness in the chemical potential, arising e.g. due to charged impurities in the system. However, to maintain the $\mathbb{Z}_2$-symmetry we
assume the local potential $\delta\mu(x)$ to be identical on the two legs. The leading term added to $H_0$ of Eq. (\ref{eq:CleanLimitHamStructure}) is of the form
\begin{align}
H_\eta & =\frac{1}{2}\sum_\nu\int dx \left(\eta(x)e^{i2 \theta_\nu}+\eta^\ast(x) e^{-i2 \theta_\nu} \right)\nonumber \\
& =\int dx  \left(\eta(x)e^{i\sqrt{2} \theta_+}+\eta^\ast(x) e^{-i\sqrt{2} \theta_+} \right)\cos{\sqrt{2}\theta_-}
\label{eq:Hdis}
\end{align}
where $\eta(x)$ is a complex random variable obeying
\begin{align}
\left<\left<\eta(x)\eta(x^\prime)\right>\right> &=\left<\left<\eta(x)\right>\right>=0 \; , \nonumber \\
\left<\left<\eta(x)\eta^\ast(x^\prime)\right>\right> &=D_\eta \delta(x-x^\prime)\; .
\label{eq:D_eta_def}
\end{align}
Distinctly from all the previous disorder terms, this introduces a non-trivial coupling term between the symmetric and antisymmetric sectors of $H_0$. In terms of their low-energy degrees of freedom, it corresponds to a many-body Boson-Fermion interaction, which in particular does not have a simple local form in terms of the Fermion fields of Eq. (\ref{eq:AntiSymmetricSector}).

Accounting for all types of disorder introduced in (a) and (b) as weak perturbations of $H_0$, we next derive RG equations in the spirit of the analysis described, e.g., in Ref. \onlinecite{GS} (see App. \ref{app:RG} for details). It is noteworthy that the special case $D_\eta=0$, which allows treatment of the $\pm$-sectors independently, indeed reduces the problem to models studied elsewhere in the literature. However, the more generic case where $D_\eta$ is finite yields a set of coupled RG equations which affects all parameters of the model, and in particular generates all other types of disorder (most prominently, $D_g$ and $D_{_-}$) even when their bare values are zero. Below we sketch the main steps
and results of this RG analysis.

\section{RG analysis and main results}
\label{sec:RGanalysis}

The various disorder terms described in the previous Section affect the behavior of the system in different ways. We use a perturbative momentum-shell RG method (see App. \ref{app:RG} for details) in order to determine their effect on the system, which will allow us to explore the different parts of the phase diagram. We shall begin with the case $D_\eta=0$ and analyse the disorder terms of type (a), which affect each of the $\pm$-sectors independently (subsections A,B below); in subsection C we introduce $D_\eta\not=0$ which couples the two sectors, and yields the full phase diagram.

\subsection{Symmetric Sector}

First, we consider the Symmetric sector $H_+$. The clean part is described by Eq. (\ref{eq:SymmetricSector}), to which we add three types of disorder - $D_U$, $D_s$ (Eq. (\ref{eq:D_Us_def})) and $D_g$ (Eq. (\ref{eq:Hdis_g})). The full Hamiltonian for the symmetric sector acquires the form
\begin{align}
H_+= & \int dx \left[\mathcal{H}_{LL}^{(+)}+\delta U(x) \left(\partial_x\theta_+\right)^2 \right.\\
+ & \left. \delta\rho_s(x) \left(\partial_x\phi_+\right)^2 + \left[\delta g(x) e^{i\sqrt{8}\theta_+}+h.c.\right]\right]\; ;
\label{eq:SymmSectorDis}
\end{align}
note that here we assume a generic filling for which the oscillatory cosine term in Eq. (\ref{eq:SymmetricSector}) can be eliminated.
The quadratic disorder parameters are better written in a dimensionless form:
\begin{equation}
\mathcal{D}_{U/s}=\frac{D_{U/s}\Lambda}{(2\pi)^4 u_{_+}^2}
\end{equation}
where $\Lambda$ is the upper momentum cutoff.
Along with the definition from Ref. \onlinecite{GS}
\begin{equation}
\mathcal{D}_g = \frac{D_g}{\Lambda^3 u_{_+}^2}
\end{equation}
one can write the RG equations for the symmetric sector:
\begin{equation}
\begin{aligned}
\frac{dK_{_+}}{d\ell}=&4\left(\mathcal{D}_s K^2_{_+} + \frac{\mathcal{D}_g}{K_{_+}}- \frac{\mathcal{D}_U}{ K^2_{_+}}\right) K_{_+} \\
\frac{du_{_+}}{d\ell}=&-4\left(\mathcal{D}_s K^2_{_+} + \frac{\mathcal{D}_g}{K_{_+}} + \frac{\mathcal{D}_U}{ K^2_{_+}}\right) u_{_+} \\
\frac{d\mathcal{D}_{s/U}}{d\ell} =& -\mathcal{D}_{s/U} \\
\frac{d\mathcal{D}_g}{d\ell} =& \left(3-\frac{4}{K_{_+}}\right)\mathcal{D}_g
\label{RG_eqn_symm}
\end{aligned}
\end{equation}
where $\ell$ is the logarithmic rescaling factor.
One readily observes that the disorder in the quadratic terms ($\mathcal{D}_{s/U}$) is always irrelevant, and so it just renormalizes the parameters $K_{_+}$ and $u_{_+}$ (see App. \ref{D_UsAnalysis}). The Luttinger parameter $K_{_+}$ can be renormalized either upwards or downwards, while the velocity is always corrected downwards -- this results from breaking the Lorenz invariance of the system. As these disorder terms only contribute corrections to the parameters of the clean model, and are never relevant, in the forthcoming more complex analysis we will ignore them and just use the effective values of $K_{_+},u_{_+}$.

On the contrary, $\mathcal{D}_g$ turns relevant at $K_c=\frac{4}{3}$. This critical value might be  modified due to $D_{s/U}$, but the general structure is the same. The exact value depends on the parameters, but around $K_{_+}=K_c$ one can find a critical manifold where the symmetric sector undergoes a SIT. The superconducting phase is a LL with power-law correlations which manifests zero resistance only in the limit $T\to0$, and the insulating phase is a disordered insulator, dominated by $D_g$.

\subsection{AntiSymmetric Sector}

\begin{figure*}
	\includegraphics[width=17cm,height=!]{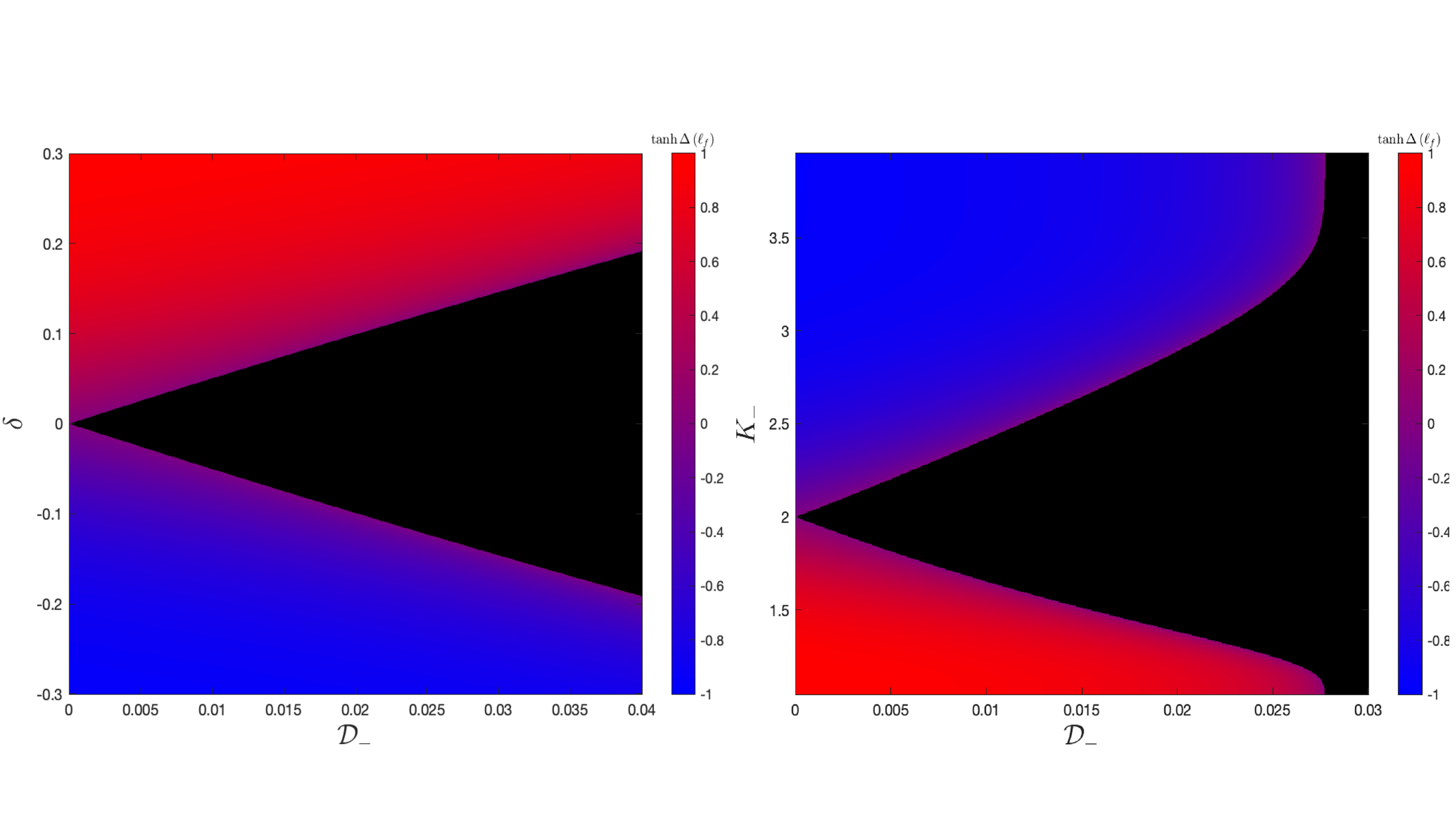}
	\caption{Left panel: Phase diagram of the anti-symmetric sector for $D_\eta=0$, as a function of the gap $\Delta$ and disorder in the gap $\mathcal{D}_-$. Here $u_{_-}=1$; the color code denotes $\Delta(\ell_f)$ obtained from the RG flow up to $\ell_f=10$.
The black region corresponds to the disordered phase where the perturbative analysis breaks down.  Right panel: the same phase diagram, parametrized by the Luttinger parameter of the antisymmetric sector $K_{_-}$; the dependence of the gap on $K_{_-}$ is monotonic but non-linear, taken from Ref. \onlinecite{Mazo2014}. In both panels, red is superconductor and blue is insulator}
\label{fig:phase_diagram_AS}
\end{figure*}

Next, we turn to the Antisymmetric sector $H_-$. The clean part of the Hamiltonian is given most generally by Eq. (\ref{eq:AntiSymmetricSDSG}). One should note that for very low values of $K_{_-}$ ($K_{_-}<1$), the term $g_\theta\cos\sqrt{8}\theta_-$ turns irrelevant and the CDW order it tends to induce is completely suppressed. In that case, the system will be a gapped superconductor, where the relative phase $\phi_{_-}$ is locked by the Josephson coupling. Similarly, for high values of $K_{_-}$ ($K_{_-}>4$), the system is a gapped insulator with CDW order parameter.

For values of $K_{_-}$ in the intermediate regime between these two extremes the system is well-described by the Fermionic Hamiltonian Eq. (\ref{eq:AntiSymmetricSector}), on which we focus. Apart from the disorder terms $D_{s/U}$, whose effects we will include in the definitions of $u_{_-}$ and $K_{_-}$, here the dominant disorder will be in the gap parameter $\Delta$ -- i.e., $D_-$ defined in Eq. (\ref{eq:D_-_def}). This disorder can be treated similarly to $D_g$, but as $D_-$ is always relevant we account for second order contributions to get a more accurate description of the behavior. We define the dimensionless disorder parameter
\begin{equation}
\mathcal{D}_-=\frac{D_-}{(2\pi)^2u_{_-}^2\Lambda};
\end{equation} 
along with the normalized gap $\delta\equiv\frac{\Delta}{u_{_-}\Lambda}$ one can write the equations:
\begin{equation}
\begin{aligned}
\frac{d\mathcal{D}_-}{d\ell}=&\mathcal{D}_-+\frac{4}{3}\frac{\delta^2}{(1+\delta^2)^2}\mathcal{D}_-^2 \\
\frac{du_{_-}}{d\ell}=&-\frac{\mathcal{D}_-}{1+\delta^2}u_{_-} \\
\frac{d\delta}{d\ell}=&\left(1-\frac{2\mathcal{D}_-}{1+\delta^2}\right)\delta\; .
\label{RG_eqn_as}
\end{aligned}
\end{equation}

The above form hints towards a normalized disorder parameter, which accounts better for the simultaneous growth of $\delta$:
$\mathcal{\tilde{D}_-}\equiv\frac{\mathcal{D}_-}{1+\delta^2}$, that obeys the equation
\begin{equation}
\frac{d\mathcal{\tilde{D}_-}}{d\ell}=\mathcal{\tilde{D}_-}\frac{1-(1-\frac{16}{3}\mathcal{\tilde{D}_-})\delta^2}{1+\delta^2}\; .
\label{RG_eqn_as_normal}
\end{equation}
This yields a threshold value $\tilde{\mathcal{D}_-}=\frac{3}{16}$ below which the disorder $\tilde{\mathcal{D}_-}$ is less relevant than $\delta$, and the system in dominated by the clean limit (see further discussion in \ref{app:DminusAnalysis}).

The resulting phase diagram is depicted in Fig. \ref{fig:phase_diagram_AS}. For low enough initial $\Delta,\mathcal{D}_-$ one finds that the disorder dominates in a triangle in parameter space, $\left|\delta\right|\leq \alpha \mathcal{D}_-$ with some constant $\alpha$. Higher initial values of $\mathcal{D}_-$ introduce non-linearity in the behavior, but the qualitative behavior is the same -- phase-locked superconductor for large positive $\Delta$ (low $K_{_-}$), disorder for small $\Delta$ (intermediate $K_{_-}$), and charge-locked insulator for large negative $\Delta$ (high $K_{_-}$). 

The primary conclusion is that here, the SIT occurs via an intermediate phase characterized by a wide distribution of the gap parameter $\Delta$, with tails in either signs. While a full characterization of its behavior requires a non-perturbative method, we interpret this phase as a Bose glass which exhibits local CDW or SC order in random locations in space (and likely developes to a Griffith's phase in the strong disorder limit). Interestingly, this $T=0$ phase diagram resembles the finite $T$ characteristic of a clean Ising transition in 1+1-dimensions, with $\mathcal{D}_-$ providing the analogue of temperature; a direct transition from CDW to SC occurs only at the singular QCP $\mathcal{D}_-=\Delta=0$.

\subsection{Disordered Coupling Term}

We next analyze the disorder term $H_\eta$ [Eq. (\ref{eq:Hdis})], which couples the $\pm$-sectors. As a basis for a perturbative RG analysis of this term, we assume $D_\eta\ll u^2_\pm\Lambda^3$ and
take advantage of the known correlations of the operators $\cos{\sqrt{2} \theta_-}$, $e^{\pm i\sqrt{2} \theta_+}$ in their respective independent unperturbed states dictated by $H_\pm$ [Eqs. (\ref{eq:SymmetricSector}), (\ref{eq:AntiSymmetricSDSG})].
Notably, the former operator controlling the coupling to the antisymmetric mode does not have a simple representation in terms of the Fermions characterizing the low-energy degrees of freedom for intermediate values of $K_{_-}$ [Eq. (\ref{eq:AntiSymmetricSector})]. However, its space-time correlations are well-characterized in terms of the order and disorder Ising fields \cite{27 Gogolin Nersesyan 98}, and are crucially dependent on the sign of $\Delta$. In particular, for $\Delta>0$ where the anti-symmetric sector is in the SC phase, its correlations are exponentially decaying.

In the insulating phase of $H_-$ established for $\Delta<0$, the disorder term $H_\eta$ couples to a more relevant operator compared to the SC phase. However, its ultimate effect on the behavior of the system depends on an additional ingredient of the model: the {\it sign} of the inter-leg interactions $V$ [see Eq. (\ref{eq:CleanLimitIntHam})], or equivalently the coefficient $g_\theta$ in Eq. (\ref{eq:AntiSymmetricSDSG}). As noted in Sec. \ref{sec:model}, in the clean limit this determines the charge ordering pattern on the rungs: for repulsive interactions ($V>0$), the charge field is locked at either one of the minima $\theta_-=\pm\nicefrac{\pi}{\sqrt{8}}$ for which $\langle\cos{\sqrt{2} \theta_-}\rangle=0$; in the case of attractive interactions ($V<0$), $\theta_-=0,\nicefrac{\pi}{\sqrt{2}}$ yielding $\langle\cos{\sqrt{2} \theta_-}\rangle\neq 0$, and hence $H_\eta$ couples to the total-charge mode via a highly relevant operator $\cos{\sqrt{2} \theta_+}$. We therefore separate these two cases in our analysis.

We first consider the repulsive interaction case $V>0$, which turns out to yield a richer phase diagram.
In this case, $\cos\sqrt{2}\theta_{-}$ has no expectation value, and its correlators decay exponentially. As a result, the only contribution of $H_\eta$ to the RG equations to leading (linear) order in $D_\eta$ will be manifested as shifts of the parameters $g$, $g_\theta$ in the clean model. The leading non-trivial contribution beyond that arises from fourth-order in the perturbation expansion; it can be interpreted as a quenched disorder term of a higher scaling dimension compared to all types of terms introduced in Sec. \ref{sec:model}, and is irrelevant in all the ordered phases of the $D_\eta=0$ case (see App. \ref{app:RG} for details). Consequently, the latter effect of a finite $D_\eta$ on the RG equations can be neglected.

To set up the derivation of RG equations at finite $D_\eta$, we first define the dimensionless disorder parameter
\begin{equation}
\mathcal{D}_\eta\equiv\frac{D_\eta}{u^2\Lambda^3};
\end{equation}
here a velocity scale $u\approx\min\left\{u_{_+},u_{_-}\right\}$ is introduced, noting that $D_\eta$ couples to both sectors.
As discussed above, the effect of $\mathcal{D}_\eta$ on the RG flow strongly depends on the behavior of the anti-symmetric mode in the clean limit. We therefore consider below three limits, classified most conveniently by the (bare) value of the parameter $K_{_-}$: the Bosonic superconductor, the Bosonic insulator, and the intermediate Fermionic regime.

{\it Bosonic superconductor ($K_{_-}<1$).} This regime is established when the last term in Eq. (\ref{eq:AntiSymmetricSDSG}) is irrelevant, and $H_-$ reduces to a standard sine-Gordon model dominated by the single cosine term describing Josephson coupling on the rungs. The anti-symmetric mode is then in a gapped phase where the relative phase field $\phi_-$ is locked at $\phi_-=0$; low-energy quantum fluctuations in $\phi_-$ are well-described by a massive Bosonic model. We note that this behavior is not significantly altered even if randomness in the mass is introduced (see App. \ref{app:OtherDis}). As already noted, in this case any operator of the form $\cos\gamma\theta_-$ coupling to the dual field is exponentially irrelevant. As a result, the sole effect of $\mathcal{D}_\eta$ is to provide corrections to the other parameters of the model which can be absorbed in their bare values, and hence practically ignored.

{\it Bosonic insulator ($K_{_-}>4$).} In this regime of parameters, the last term in Eq. (\ref{eq:AntiSymmetricSDSG}) is dominant while the Josephson coupling on the rungs turns irrelevant. As a result, one obtains a strong tendency for charge-locking in the anti-symmetric sector at a CDW pattern obeying $\langle\cos\sqrt{8}\theta_-\rangle\approx -1$, and $H_-$ can be approximated by a massive Bosonic model with gap $|\Delta|$ (in terms of the definitions of Sec. \ref{sec:model}, $\Delta<0$). However, since at the same time $\langle\cos\sqrt{2}\theta_-\rangle\approx 0$, the leading contribution to the RG equation for the disorder term arises from order $D_\eta^2$ (see App. \ref{app:RG}).
The linear order in $\mathcal{D}_\eta$, on the other hand, generates terms which can be regarded as corrections to the various parameters of $H_0$. Combining them all, we get the following set of coupled equations:
\begin{equation}
\begin{aligned}
\frac{dK_{_+}}{d\ell}  = & 4(\mathcal{D}_sK_{_+}^2+\frac{2\mathcal{D}_g}{K_{_+}}-\frac{\mathcal{D}_U}{K_{_+}^2})K_{_+} \\
& + \left(\frac{u_{_+}}{u}\right)\mathcal{D}_\eta \\
\frac{du_{_+}}{d\ell} = & -4(\mathcal{D}_sK_{_+}^2+\frac{2\mathcal{D}_g}{u_{_+}}+\frac{\mathcal{D}_U}{K_{_+}^2})u_{_+} \\
& - \left(\frac{u_{_+}}{u}\right)\frac{u_{_+}}{K_{_+}} \mathcal{D}_\eta \\
\frac{dK_{_-}}{d\ell} = & 4(\mathcal{D}_sK_{_-}^2-\frac{\mathcal{D}_U}{K_{_-}^2})\frac{K_{_-}}{1+\delta^2}\\
& + \left(\frac{u_{_-}}{u}\right)\mathcal{D}_\eta \\
\frac{du_{_-}}{d\ell} = & -4(\mathcal{D}_sK_{_-}^2+\frac{\mathcal{D}_U}{K_{_-}^2})\frac{u_{_-}}{1+\delta^2} \\
& - \left(\frac{u_{_-}}{u}\right) \frac{u_{_-}}{K_{_-}} \mathcal{D}_\eta \\
\frac{d\delta}{d\ell} = & \delta - \delta^{\nicefrac{2}{K_{_-}}-1}\frac{u_{_+}}{u_{_-}} \mathcal{D}_\eta \\
\frac{d\mathcal{D}_U}{d\ell} = & -\mathcal{D}_U  \\
\frac{d\mathcal{D}_s}{d\ell} = & -\mathcal{D}_s  \\
\frac{d\mathcal{D}_g}{d\ell} = & \left(3-\frac{4}{K_{_+}}\right) \mathcal{D}_g  + c\mathcal{D}_\eta^2\\
 \frac{d\mathcal{D}_\eta}{d\ell} = & \left(\frac{3}{2}-\frac{2}{K_{_+}}-\frac{2}{K_{_-}\left(1+\delta^2\right)}\right)\mathcal{D}_\eta
\label{RG_BosonI}
\end{aligned}
\end{equation}
where $c$ is a constant of order unity.

It is noteworthy that the two Bosonic descriptions mentioned above are valid approximations even when both cosine terms are relevant, if one of them has a significantly larger effect on the system, as quantified by the gaps they induce \cite{Mazo2014}; see App. \ref{app:H_0}.

{\it Fermionic regime (intermediate values of $K_{_-}$).} In this regime where the clean part of the anti-symmetric sector is best approximated by the Fermionic model Eq. (\ref{eq:AntiSymmetricSector}), the effect of $\mathcal{D}_\eta$ on the RG equations is similar in nature to the previous case; the primary difference is that the operator $\cos\sqrt{8}\theta_-$ which couples to the leading terms generated by the disorder can be more conveniently expressed in terms of Fermion fields. This yields the following set of coupled equations (see App. \ref{app:RG} for details):
\begin{equation}
\begin{aligned}
\frac{dK_{_+}}{d\ell} = & 4(\mathcal{D}_sK_{_+}^2+\frac{2\mathcal{D}_g}{K_{_+}}-\frac{\mathcal{D}_U}{K_{_+}^2})K_{_+} \\
& + \left(\frac{u_{_+}}{u}\right) \mathcal{D}_\eta \\
\frac{du_{_+}}{d\ell} = & -4(\mathcal{D}_sK_{_+}^2+\frac{2\mathcal{D}_g}{K_{_+}}+\frac{\mathcal{D}_U}{K_{_+}^2})u_{_+} \\
& - \left(\frac{u_{_+}}{u}\right) \frac{u_{_+}}{K_{_+}} \mathcal{D}_\eta \\
\frac{du_{_-}}{d\ell} = & -4(\mathcal{D}_sK_{_-}^2+\frac{\mathcal{D}_U}{K_{_-}^2})\frac{u_{_-}}{1+\delta^2} \\
& - \left(\frac{u_{_-}}{u}\right) \frac{u_{_-}}{K_{_-}} \mathcal{D}_\eta - \frac{\mathcal{D}_-}{1+\delta^2}u_{_-}\\
\frac{d\delta}{d\ell} = & \delta - C_{g_\theta} \frac{u_{_+}}{u_{_-}} \mathcal{D}_\eta - \frac{2\mathcal{D}_-}{1+\delta^2}\delta \\
\frac{d\mathcal{D}_U}{d\ell} = & -\mathcal{D}_U \\
\frac{d\mathcal{D}_s}{d\ell} = & -\mathcal{D}_s \\
\frac{d\mathcal{D}_g}{d\ell} = & \left(3-\frac{4}{K_{_+}}\right) \mathcal{D}_g + c_1 \mathcal{D}_\eta^2 \\
\frac{d\mathcal{D}_-}{d\ell} = & \mathcal{D}_- + \frac{4}{3} \frac{\delta^2}{(1+\delta^2)^2} \mathcal{D}_-^2 + c_2 \mathcal{D}_\eta^2\\
 \frac{d\mathcal{D}_\eta}{d\ell} = & \left(\frac{1}{2}-\frac{2}{K_{_+}}\right)\mathcal{D}_\eta \; .
\label{RG_Fermion}
\end{aligned}
\end{equation}
Here $c_1$, $c_2$ are constants of order unity; $C_{g_\theta}=\left(\frac{16\pi g_\theta}{u_{_-}K_{_-}\Lambda^2}\right)^{\frac{\nicefrac{2}{K_{_-}}-1}{2-\nicefrac{2}{K_{_-}}}}$ does not change significantly in the regime of parameters where Eq. (\ref{RG_Fermion}) is valid, so one can consider it as a constant as well.

It is evident from the above two sets of equations that
in both cases, $\mathcal{D}_\eta$ turns relevant for high values $K_{_+}$ which exceed the critical point ($K_{_+}=\frac{4}{3}$) for $\mathcal{D}_g$ to become relevant. Beyond this critical point which indicates a localization transition in the symmetric sector, the perturbative analysis breaks down, leading to a rapid growth of $K_{_+}$ and consequently of $\mathcal{D}_\eta$.
We therefore conclude that there is effectively a unique disordered insulating phase. Within the framework of the weak-disorder approximation, it is not possible to infer the precise nature of the charge-density pattern on the rungs in this phase, though it may survive locally in randomly distributed disconnected domains.

It should be noted, however, that while $\mathcal{D}_\eta$ does not tune a phase-transition separable from the one dominated by $\mathcal{D}_g$, its coupling to both the symmetric and anti-symmetric sectors generates a flow of all the other parameters [see Eqs. (\ref{RG_BosonI}), (\ref{RG_Fermion})]. As a result, it can serve as the tuning parameter for various transitions, as can be seen in Fig. \ref{fig:phase_diagram}.
This figure was obtained by setting the bare parameters to the Fermionic regime where the RG flow is determined by Eq. (\ref{RG_Fermion}), and exhibits a pronounced effect of $\mathcal{D}_\eta$. We identify four distinct phases, accessible e.g. by tuning $\mathcal{D}_\eta$ upwards: for relatively low values of $\mathcal{D}_\eta$ and $K$, the symmetric mode is a gapless LL while the anti-symmetric mode undergoes a transition from a phase-locked SC phase to a CDW-ordered insulator via a disordered intermediate phase, whose nature is described in subsection B above; the fourth phase realized beyond a critical line in the $\mathcal{D}_\eta$--$K$ plane is a disordered insulator, characterized primarily by localization of the symmetric charge mode. Since $D_\eta$ couples the sectors, this will be the case in the antisymmetric sector as well.

\begin{figure*}
	\includegraphics[width=17cm,height=!]{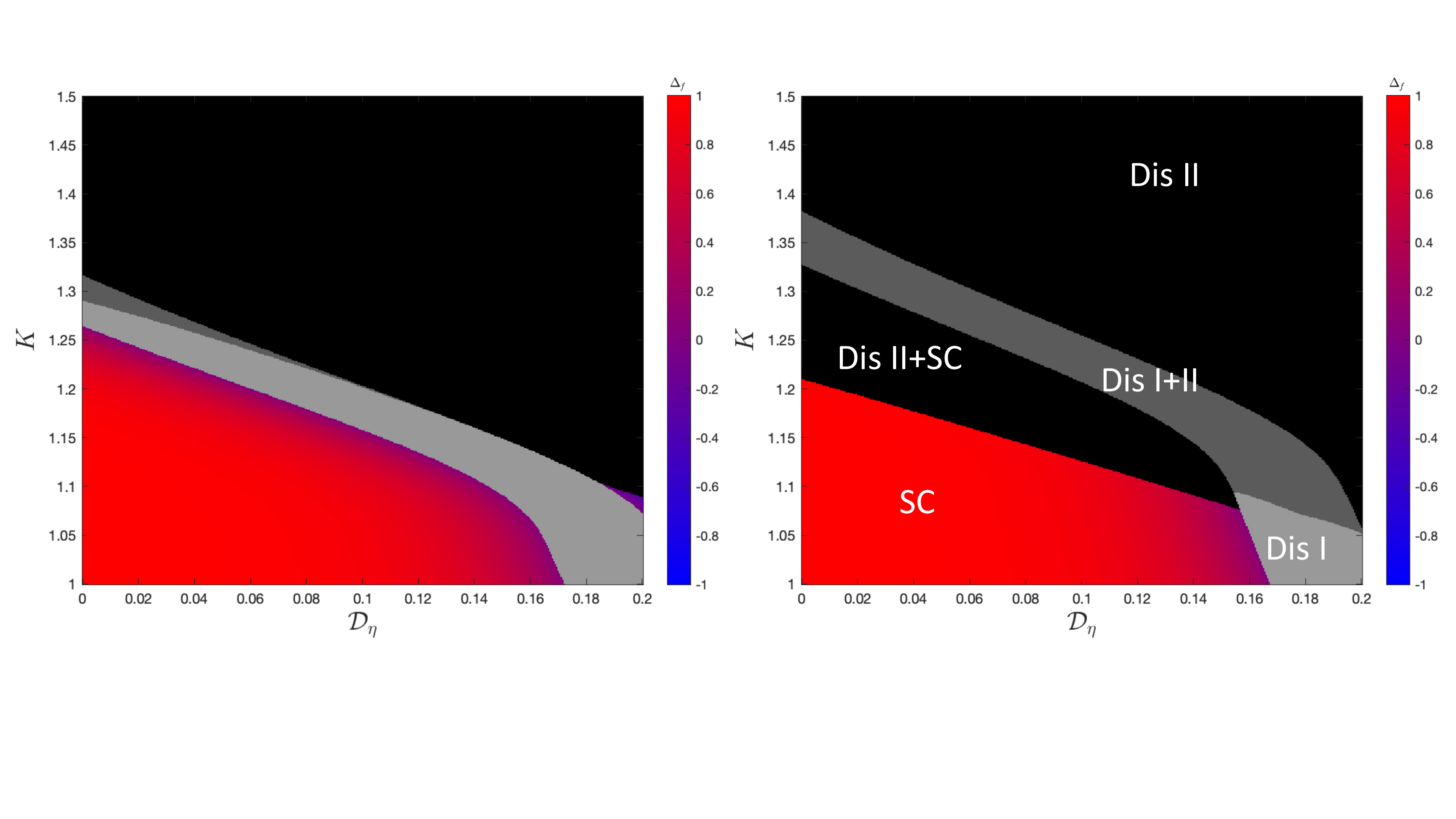}

	\caption{Phase diagrams as a function of $K$ and $\mathcal{D}_\eta$ for $V>0$, and different values of the ratio $\nicefrac{K_{_+}}{K_{_-}}$. In both panels $u_+=u_-=1$, $\mathcal{D}_g=\mathcal{D}_-=0.002$, $g_\theta=0.2$, $g_\phi=0.05$ and $\mathcal{D}_{U,s}=0$. Left: $\nicefrac{K_{_+}}{K_{_-}}=0.98$; Right:$\nicefrac{K_{_+}}{K_{_-}}=1.08$. The titles of the phases match the definitions from figure \ref{fig:phase_diagram}.}
\label{fig:phase_diagram_disordered_phases}
\end{figure*}

Although Figure \ref{fig:phase_diagram} captures the richness of the phase diagram for typical parameters, tuning the parameters differently can introduce other phases. Specifically, the tuning parameter $\frac{K_{_+}}{K_{_-}}$ can change the order of the transitions: $D_g$ may turn relevant before $D_\Delta$. Two exemplary figures with different topology of the phases diagram can be seen in Fig. \ref{fig:phase_diagram_disordered_phases}. It is suggestive that a variety of distinct disordered phases are generated (see black regions and different shades of grey in the figure).

Based on the weak-disorder approximation applied in our study, one can not reliably deduce the exact nature of these phases. However, it appears that within the regime where a disordered insulator is established in the symmetric sector, some of the independent behavior of the anti-symmetric sector still persists. In particular, there appears to be a regime where $D_\Delta$ is relevant and dominates the anti-symmetric sector, denoted by "Dis I+II" in Fig. \ref{fig:phase_diagram_disordered_phases}; more interestingly, there is potentially a mixed phase where the antisymmetric sector still exhibits robust superconductivity, denoted "Dis II+SC" in the figure.


We finally consider the crucially different case where the interactions on the rungs of the ladder are attractive, $V<0$. The most significant effect of this change of sign is manifested in the sign-reversal of the parameter $g_\theta$ in Eq. (\ref{eq:AntiSymmetricSDSG}); it is therefore equivalent to performing a shift $\theta_-\to\theta_-+\frac{\pi}{\sqrt{8}}$ in the last cosine term in $H_-$ while maintaining the other parts of the clean Hamiltonian the same. The resulting effect on the behavior of $\mathcal{D}_\eta$ is dramatic: the CDW pattern in the insulating phase of the antisymmetric sector favors $\theta_-=0$, which corresponds to equal charge densities on the two legs of the ladder. In this ground state, the operator $\cos{\sqrt{2}\theta_-}$ in $H_\eta$ [Eq. (\ref{eq:Hdis})] has a finite expectation value.

As a result, $\mathcal{D}_\eta$ has a much larger contribution, and in the charge-locked phase of the anti-symmetric mode yields the following RG equation:
\begin{equation}
\frac{d\mathcal{D}_\eta}{d\ell} = (3-\frac{1}{K_{_+}})\mathcal{D}_\eta \; .
\end{equation}
This corresponds to a highly relevant disorder, with critical Luttinger parameter of $K_{+c}=\frac{1}{3}$. We conclude that once the antisymmetric sector transitions into the insulating phase (which happens for $K_{_-}$ way above this critical value), $\mathcal{D}_\eta$ turns relevant. Notably, since its effect on the other disorder parameters is similar to what we have obtained in the $V>0$ case, this induces their divergence and consequently the formation of a disordered insulating phase. The prominent change in the phase diagram as compared to Fig. \ref{fig:phase_diagram} is that the intermediate phase manifesting CDW order on the rungs will disappear; the sole intermediate phase separating the SC from a disordered insulator will be the mixed phase marked by a grey region in Fig. \ref{fig:phase_diagram}, where randomness in established in the anti-symmetric sector while the symmetric mode remains gapless.

\section{Discussion}
\label{sec:discussion}

In this work we have discussed a two-leg ladder model of a strongly fluctuating superconductor in the presence of various types of quenched disorder, and examined the resulting $T=0$ phases. We found that by tuning a parameter $K$ -- which is controlled by the ratio of charging and Josephson energies -- or the disorder strength, the system typically undergoes a sequence of quantum phase transitions rather than a direct SIT. Between the two extreme phases -- a SC phase manifesting robust phase-locking on the rungs, and a disordered insulator -- at least one intermediate phase is formed. Particularly, in the case of repulsive interactions across the rungs, two distinct intermediate phases are identified: one ordered and one disordered. The ordered phase is characterized by a dipolar CDW order on the rungs, while the disordered intermediate phase exhibits Grifith's singularities interpolating between phase and charge locking on the rungs; in both these mixed phases, as well as in the SC phase, the longitudinal plasmon mode (corresponding to fluctuations in the total charge) maintains a gapless LL behavior and perfect conduction ($R\rightarrow 0$) is achieved in the limit $T=0$. At the opposite extreme, a full-fledged localization of this mode occurs only in the disordered insulator phase (see Fig. \ref{fig:phase_diagram}).

The richness of this phase diagram rests on the perfect symmetry between the two legs, which restricts spatial variations in the parameters to the longitudinal direction. This ensures a relative resilience to weak disorder of the separability between anti-symmetric (transverse) and symmetric (longitudinal) phase/charge fluctuation modes. As a consequence, in the former sector we observe traces of behavior characteristic to the Ising model in a random transverse field \cite{randomIsing}. Most prominently, in the case where the inter-leg charge interaction is repulsive, the system supports the two (duality-related) gapped phases reminiscent of the clean Ising model, separated by a mixed phase where segments of phase-locked rungs are embedded in a background of charge-locked rungs, or vice versa.

Our findings suggest that engineered Josephson ladders in the strong quantum-fluctuations regime ($E_C\sim E_J$) can serve as a compelling platform for simulating the physics of the random-field Ising model, as well as coupled Boson-Fermion systems in 1D -- as long as they possess the above mentioned $\mathbb{Z}_2$-symmetry. Moreover, provided separate contacts to the edges of the two legs are accessible, probing of the various phases is possible via measurement of different conductance components \cite{Mazo2014}. In practice, however, accurate control of the parameters in Josephson arrays which ensures the perfect symmetry is quite challenging. An alternative realization of the model where the discrete degree of freedom is internal rather than a spatially separate leg (e.g., a spin, valley or orbital index) can possibly be designed, e.g. in cold atom systems or in van der Waals materials with long-range disorder.


\label{sec:discussion}

\acknowledgements
We gratefully acknowledge useful discussions with Ehud Altman, Ganpathy Murthy, Jonathan Ruhman and Thomas Vojta. E. S. thanks the Aspen Center for Physics (NSF Grant No. 1066293) for its hospitality. This work was supported by the US-Israel Binational Science Foundation through awards No.
2016130 and 2018726, and by the Israel Science Foundation (ISF) Grant No. 993/19.

\appendix

\section{Derivation of the low-energy theory}
\label{app:H_0}

The ladder model in the clean limit is described in Eqs. (\ref{eq:CleanLimitHamStructure})-(\ref{eq:CleanLimitIntHam}). In this Appendix we detail the derivation of its low-energy approximation, Eqs. (\ref{eq:SymmetricSector})--(\ref{eq:AntiSymmetricSector}), used as a basis for the remains of the analysis.

As a first stage, in the definition of $\rho_\nu$ in terms of $\theta_\nu$ [Eq. (\ref{eq:rho_nu})] we keep only the leading harmonics, coming from $p=\pm 1$. The Hamiltonian acquires the structure
\begin{widetext}
\begin{equation}
\begin{aligned}
\mathcal{H}_\nu=&\frac{U}{2\pi^2} \left(\partial_x\theta_\nu\right)^2+\frac{\rho_s}{2}\left(\partial_x\phi_\nu\right)-\frac{2\rho_0U}{\pi}\partial_x\theta_\nu\cos{\left(2\left(\theta_\nu-\pi\rho_0x\right)\right)}+2U\rho_0^2\cos{\left(2\left(\theta_\nu-\pi\rho_0x\right)\right)}^2 \\
\mathcal{H}_{int}=&\frac{V}{\pi^2}\partial_x\theta_1\partial_x\theta_2-\frac{2\rho_0V}{\pi}\left[\partial_x\theta_1\cos{\left(2\left(\theta_2-\pi\rho_0x\right)\right)}+\partial_x\theta_2\cos{\left(2\left(\theta_1-\pi\rho_0x\right)\right)}\right] \\
+&4\rho_0^2V\cos{\left(2\left(\theta_1-\pi\rho_0x\right)\right)}\cos{\left(2\left(\theta_2-\pi\rho_0x\right)\right)}\; .
\label{eq:H12}
\end{aligned}
\end{equation}
\end{widetext}
The terms combining gradients with cosines must vanish, as they are not symmetric to inversion ($x\mapsto-x$). We re-write Eq. (\ref{eq:H12}) in terms of $\theta_\pm,\phi_\pm$ as defined in Eq. (\ref{eq:SectorDefinition}), describing the symmetric ($+$) and antisymmetric ($-$) sectors to get the following Hamiltonian:
\begin{widetext}
\begin{equation}
\begin{aligned}
H=\int dx&\left[ \frac{U+V}{2\pi^2}\left(\partial_x\theta_+\right)^2+\frac{\rho_s}{2}\left(\partial_x\phi_+\right)^2+2\rho_0^2V \cos\left(\sqrt{8}\theta_+-4\pi\rho_0x\right) \right.\\
+&\frac{U-V}{2}\left(\partial_x\theta_-\right)^2+\frac{\rho_s}{2}\left(\partial_x\phi_-\right)^2+2\rho_0^2V \cos\left(\sqrt{8}\theta_-\right)-J\cos\left(\sqrt{2}\phi_-\right) \\
+&\left.U\rho_0^2\cos\left(\sqrt{8}\theta_++\sqrt{8}\theta_--4\pi\rho_0x\right)+U\rho_0^2\cos\left(\sqrt{8}\theta_+
-\sqrt{8}\theta_--4\pi\rho_0x\right)\right]\; .
\label{eq:Hpm}
\end{aligned}
\end{equation}
\end{widetext}
The first two lines correspond to Eqs. (\ref{eq:SymmetricSector}) and (\ref{eq:AntiSymmetricSDSG}) in the main text. The terms on the third line are less relevant, being higher harmonics which we already neglect in the density operator $\rho_\nu$.

We now identify the quadratic part of each sector as a Luttinger Liquid, with $u_{_\pm}=\sqrt{\left(U\pm V\right)\rho_s}$ and $K_{_\pm}=\sqrt{\frac{U\pm V}{\pi^2\rho_s}}$. We see that for the bare values, $K_{_+}>K_{_-}$ for repulsive interactions, and $K_{_+}<K_{_-}$ for attractive interactions. However, as we show in App. \ref{app:D_UsAnalysis}, disorder terms of the type $\mathcal{D}_{U/s}$ modify each of these parameters independently; therefore, this hierarchy of the Luttinger parameters is not necessarily maintained once disordedr is introduced.

To further analyze the antisymmetric sector, a slight modification of the standard Fermionization \cite{27 Giamarchi 04} is helpful:
\begin{equation}
\psi_r=\frac{U_r}{\sqrt{2\pi a}} e^{-i\left[\nicefrac{r\phi_-\left(x\right)}{\sqrt{2}}-\sqrt{2}\theta_-\left(x\right)\right]}
\end{equation}
with $r=R,L$ for right- and left-moving Fermions, $a$ the lattice constant, and $U_r$ the Klein factor. For $K_{_-}=2$, one can exactly map Eq. (\ref{eq:AntiSymmetricSDSG}) to non-interacting Fermions:
\begin{equation}
\begin{aligned}
&\psi_R^\dagger\left(-i\partial_x\right)\psi_R-\psi_L^\dagger\left(-i\partial_x\right)\psi_L \\
&=\frac{2\left(\partial_x\theta_-\right)^2+\frac{\left(\partial_x\phi_-\right)^2}{2}}{\pi} \\
\psi_R^\dagger\psi_L+h.c.&=\frac{1}{\pi a} \cos\left(\sqrt{2}\phi_-\right) \\
\psi_L\psi_R+h.c.&=\frac{1}{\pi a} \cos\left(\sqrt{8}\theta_-\right)
\end{aligned}
\end{equation}

Now one can decompose these Fermions to Majorana (real) fields
\begin{equation}
\psi_r=\frac{\xi_{r1}+i\xi_{r2}}{\sqrt{2}}
\end{equation}
and the Hamiltonian decouples into two independent sectors: one with $\xi_{\uparrow R}\equiv\xi_{1 R}$ and $\xi_{\uparrow L}\equiv\xi_{2 L}$, and the other with $\xi_{\downarrow R}\equiv\xi_{2 R}$ and $\xi_{\downarrow L}\equiv\xi_{1 L}$. The Hamiltonian in terms of these Majorana fields is
\begin{equation}
\begin{aligned}
H=&\sum_{\nu=\updownarrow}\int dx u_{_-}\left[\xi_{R\nu}\left(-i\partial_x\right)\xi_{R\nu}-\xi_{L\nu}\left(-i\partial_x\right)\xi_{L\nu}\right] \\
&-i\Delta_\nu\xi_{R\nu}\xi_{L\nu}\; .
\end{aligned}
\end{equation}

In the case of $K_{_-}=2$, $\Delta_\updownarrow=\Delta_\theta\pm\Delta_\phi$ where $\Delta_{\theta,\phi}$ are linear in $g_{\theta,\phi}$, the coefficients of the cosines in Eq. (\ref{eq:AntiSymmetricSDSG}). However, if $K_{_-}\neq2$, there is an interaction term $\propto\left(K_{_-}-2\right)\xi_{\uparrow R}\xi_{\uparrow L}\xi_{\downarrow R}\xi_{\downarrow L}$. Provided there is a separation of energy scales between the $\updownarrow$-sectors, this can be treated in mean-field; the effective gaps have the same structure, but with $\Delta_{\theta,\phi}$ the gap of the corresponding sine-Gordon model.
This approximation is self-consistent if $\Delta_\theta\approx\Delta_\phi$ ($\Delta_\downarrow\ll\Delta_\uparrow$), in which case integrating over the $\uparrow$ sector is justified and yields the low-energy theory. In this case, the effective Hamiltonian is dominated by the sector with the smaller gap $\Delta_\downarrow$ [Eq. (\ref{eq:AntiSymmetricSector}) in the main text where we have dropped the subscript $\downarrow$ on $\Delta$], which undergoes a transition as $\Delta_\downarrow$ changes sign.
This allows us to analyze the behavior of the transition in the anti-symmetric sector and identify it as an Ising transition \cite{Mazo2014}. Note that $\Delta_\uparrow$ and $\Delta_\downarrow$ change their classification when the interaction term $g_\theta\propto V$ changes its sign; we define them such that $\left|\Delta_\uparrow\right|=\left|\Delta_\theta\right|+\left|\Delta_\phi\right|$ and so $\Delta_\uparrow$ is always the larger in magnitude.

We next derive the disorder term $H_\eta$ originating from randomness in the chemical potential $\mu(x)$. Assuming a perfect symmetry between the legs and employing the leading harmonics in the expansion Eq. (\ref{eq:rho_nu}), the coupling to $\mu(x)$ is given by
\begin{equation}
\begin{aligned}
H_\mu=&-\int dx \mu\left(x\right) \left[\rho_1\left(x\right)+\rho_2\left(x\right)\right] \\
=&\int dx \mu\left(x\right) \left[\frac{1}{\pi}\left(\partial_x\theta_1+\partial_x\theta_2\right) \right.\\
-&\left.2\rho_0\left(\cos\left(2\theta_1-2\pi\rho_0 x\right)-\cos\left(2\theta_2-2\pi\rho_0 x\right)\right)\right] \\
=&\int dx \frac{\sqrt{2}\mu\left(x\right)}{\pi}\partial_x\theta_+ \\
-&\int dx 4\rho_0\mu\left(x\right)\cos\left(\sqrt{2}\theta_+-2\pi\rho_0x\right)\cos\left(\sqrt{2}\theta_-\right)
\end{aligned}
\label{eq:H_mu}
\end{equation}
The first term can be "gauged out" with the shift by a random phase:
\begin{equation}
\begin{aligned}
\theta_+ &\mapsto\theta_+ +\varphi(x)\; ,\\
\varphi(x)&\equiv \frac {\sqrt{2}\pi}{\left(U+V\right)} \int^x \mu\left(x\right)
\end{aligned}
\end{equation}
which yields
\begin{equation}
\begin{aligned}
&\cos\left(\sqrt{8}\theta_+-4\pi\rho_0x\right) \\
\mapsto & \;\frac{1}{2}e^{i\left(\sqrt{8}\theta_+-4\pi\rho_0x+\sqrt{8}\varphi\left(x\right)\right)}+h.c.\; .
\end{aligned}
\end{equation}
Substituting in the cosine of Eq. (\ref{eq:SymmetricSector}), we obtain
\begin{equation}
2\rho_0^2Ve^{i\left(\sqrt{8}\varphi\left(x\right)-4\pi\rho_0x\right)}\equiv\delta g\left(x\right)
\end{equation}
where the correlations of $\delta g$ are approximated to be totally non-correlated. Note that the amplitude $g$ might also change because of randomness in $V$.

The second term of $H_\mu$ [Eq. (\ref{eq:H_mu})] is different. With the random phase from the first term, we can write it as
\begin{equation}
\begin{aligned}
H_\eta=&\int dx \left[ \eta\left(x\right) e^{i\sqrt{2}\theta_+}+h.c.\right]\cos\left(\sqrt{2}\theta_-\right)\; ,\\
\eta\left(x\right)\equiv&4\rho_0\mu\left(x\right)e^{i\left(\sqrt{2}\varphi\left(x\right)-2\pi\rho_0x\right)}\; .
\end{aligned}
\end{equation}
Here $\eta(x)$ has both random amplitude and random phase, so we once again approximate it to be totally non-correlated.


\section{Derivation of the RG Equations}
\label{app:RG}
In this Appendix we will discuss the method used to derive the RG equations in this work, presented in Sec. \ref{sec:RGanalysis}. We particularly focus on the contribution of the disorder term in chemical potential ($\mathcal{D}_\eta$) in different regimes of the parameter space, primarily on the regime where the antisymmetric sector is well-described by Majorana fields.

Generally, we consider a disorder term with the operator $\hat{O}\left(x,\tau\right)$ of the form
\begin{equation}
S_{dis}=\int dx d\tau \left[\delta g_O\left(x\right) \hat{O}\left(x,\tau\right) \right]
\end{equation}
in which the random coefficient $\delta g_O\left(x\right)$ obeys $\langle\langle \delta g_O \rangle\rangle=0$ and the short-range correlations:
\begin{equation}
\langle\langle \delta g_O\left(x\right)\delta g_O\left(x^\prime\right)\rangle\rangle=D_O\delta\left(x-x^\prime\right)
\label{eq:DeltaCorrelations}
\end{equation}
where $\langle\langle\cdots\rangle\rangle$ stands for statistical averaging over the realizations of the disorder. We substitute this as a term in the action, write the expression for the partition function, and expand to second order in $S_{dis}$. Averaging over disorder and using Eq. (\ref{eq:DeltaCorrelations}) leads to the main contribution:
\begin{equation}
\langle\langle S_{dis}^2\rangle\rangle = \int dx d\tau d\tau^\prime D_O \hat{O}\left(x,\tau\right)\hat{O}\left(x,\tau^\prime\right)
\label{eq:BasicSdis2}
\end{equation}
and one should subtract the disconnected terms, resulting from $\langle\langle S_{dis}\rangle\rangle^2$.

To derive RG equations, we write the action describing our model in momentum space. Now we would like to apply momentum-sell renormalization group, so we begin by introducing a high momentum cutoff $\Lambda$ and splitting the fields representing the free part of the action to their slow and fast momentum components:
\begin{equation}
\begin{aligned}
\xi\left(r\right)=&\xi^<\left(r\right) + \xi^>\left(r\right) \\
\xi^<\left(r\right)=&\frac{1}{\sqrt{L\beta}}\sum_{\left|\left|q\right|\right|<\Lambda^\prime} e^{iq\cdot r} \xi\left(q\right) \\
\xi^>\left(r\right)=&\frac{1}{\sqrt{L\beta}}\sum_{\Lambda^\prime<\left|\left|q\right|\right|<\Lambda} e^{iq\cdot r} \xi\left(q\right)
\label{eq:FastSlowDecomp}
\end{aligned}
\end{equation}
where $\xi$ stands for either of the Majorana fields $\xi_R,\xi_L$ or the Bosonic fields $\phi_+,\theta_+$; here $\Lambda^\prime=e^{-d\ell}\Lambda$ is a slightly smaller momentum cutoff, $q=\left(\frac{\omega}{u},k\right)$, $r=\left(u\tau,x\right)$, and $\left|\left|q\right|\right|^2=\frac{\omega^2}{u^2}+k^2$ with the appropriate velocity $u$; $L$ and $\beta$ are respectively the length of the system and the inverse temperature.

If the operator $\hat{O}\left(x,\tau\right)$ can be directly written in terms of the fields in the free action, in our case $\theta_+,\phi_+,\xi_R,\xi_L$, one can integrate over the fast modes ($\theta_+^>,\phi_+^>,\xi_R^>,\xi_L^>$) to get an effective expression for the action of the slow modes. Below we demonstrate how this procedure can be implemented, and show that this effective expression can be cast in the following form:
\begin{equation}
\begin{aligned}
&\int dx d\tau d\tau^\prime e^{\gamma d\ell} D_O \hat{O}^<\left(x,\tau\right)\hat{O}^<\left(x,\tau^\prime\right)\\
+&\sum_i \int dx d\tau \alpha_i d\ell \hat{O}_i^<\left(x,\tau\right)
\label{eq:RGStructure}
\end{aligned}
\end{equation}
where $\left\{\hat{O}_i\right\}$ is a set of local operators and $\alpha_i$ are coefficients proportional to $D_O$; the exponent $\gamma$ is related to the scaling dimension of $\hat{O}$; finally, $\hat{O}^<$ is just $\hat{O}$ with all the fields replaced with their "slow" low-momentum component.

This effective action is defined to fulfill the following equation:
\begin{equation}
e^{-S^<_{eff}} = \int \mathcal{D}\xi^> e^{-S}
\label{eq:Seff}
\end{equation}
where we integrate over fast modes of all fields. To obtain $S_{eff}^<$, we expand the exponent around the quadratic part of the action $S_0$, which results with a perturbative description of the way $D_O$ scales and its effects on the other parameters of the model.

We begin with a simple example, the case of $\Delta$ disorder [Eq. (\ref{eq:AntiSymmetricSector}), (\ref{eq:D_-_def})], where $\delta g_O\left(x\right)=\delta\Delta\left(x\right)$ and $\hat{O}=i\xi_R\xi_L$.
The disorder-averaged expression for $S_{dis}^2$ is then given by

\vspace{1cm}

\begin{widetext}
\begin{equation}
\begin{aligned}
\langle\langle S_{dis}^2\rangle\rangle=&-D_\Delta\int dxd\tau d\tau^\prime \xi_R\left(x,\tau\right)\xi_L\left(x,\tau\right)\xi_R\left(x,\tau^\prime\right)\xi_L\left(x,\tau^\prime\right) \\
=&-D_\Delta\int d^3kd^2\omega \xi_R\left(k_1,\omega_1\right)\xi_L\left(k_2,-\omega_1\right)\xi_R\left(k_3,\omega_3\right)\xi_L\left(-k_1-k_2-k_3,-\omega_3\right)\; .
\end{aligned}
\end{equation}
\end{widetext}
In momentum space we split the 5-dimensional integral to different regimes according to the decomposition in Eq. (\ref{eq:FastSlowDecomp}). That means the integration regime is split into sixteen different parts, as each momentum vector can be in the smaller ball ($\left|q\right|<\Lambda^\prime$, "slow") or on the momentum shell ($\Lambda^\prime<\left|q\right|<\Lambda$, "fast"). However, expectation values over an odd number of fields vanish, which means a large part of the terms cancel. Among those remaining we can use some symmetries, and essentially get the following expression:
\begin{widetext}
\begin{equation}
\begin{aligned}
\langle\langle\langle S^2_\Delta\rangle_>\rangle\rangle =& - L\beta D_\Delta\int d^5 q \xi_R\xi_L\xi_R\xi_L - 2\beta D_\Delta \int dkd\omega \xi_R\left(k,\omega\right)\xi_L\left(-k,-\omega\right)\oint dk^\prime d\omega^\prime \left\langle \xi_R\left(k^\prime,\omega^\prime\right)\xi_L\left(-k^\prime,-\omega^\prime\right)\right\rangle_> \\
-&2D_\Delta \int dkd\omega \xi_R\left(k,\omega\right)\xi_L\left(-k,-\omega\right) \oint dk^\prime \left\langle \xi_L\left(k^\prime,-\omega\right)\xi_R\left(-k^\prime,\omega\right) \right\rangle_> \\
+&D_\Delta \int dk d\omega \xi_R\left(k,\omega\right)\xi_R\left(-k,-\omega\right) \oint dk^\prime \left\langle \xi_L\left(k^\prime,\omega\right)\xi_L\left(-k^\prime,-\omega\right) \right\rangle_> \\
+&D_\Delta \int dk d\omega \xi_L\left(k,\omega\right)\xi_L\left(-k,-\omega\right) \oint dk^\prime \left\langle \xi_R\left(k^\prime,\omega\right)\xi_R\left(-k^\prime,-\omega\right) \right\rangle_> \\
-L\beta &\oint d^5 q \left\langle\xi_R\xi_L\xi_R\xi_L\right\rangle_>
\label{eq:SDel2}
\end{aligned}
\end{equation}
\end{widetext}
where $d^5q=dk_1dk_2dk_3d\omega_1d\omega_3$, and the integrals $\int dk$ and $\int d\omega$ are over the smaller momentum ball $\left|q_i\right|<\Lambda^\prime$, while those denoted by $\oint dk d\omega^\prime$ are over the shell; $\oint dk^\prime$ means that $\left(k^\prime,\omega\right)$ should be on the momentum shell. Also, note we have used the fact that $\left\langle\xi\left(q\right)\xi\left(-q^\prime\right)\right\rangle = f\left(q\right) \delta_{q,q^\prime}$.

Among the resulting six terms, the first one will give us the rescaling of $D_\Delta$, the next four will be corrections to local terms like the second line of Eq. (\ref{eq:RGStructure}), and the last one is a non-interesting constant.
The expectation values are all over fast modes and with respect to the quadratic action $S_-$ (the anti-symmetric part of $S_0$). They are known, and using the approximation $\nicefrac{\omega}{u_-},k\ll\Lambda$ all the integrals are quite simple as well.

We now note that Eq. (\ref{eq:SDel2}) yields the desired correction to $S_{eff}^<$ [Eq. (\ref{eq:Seff})] only after re-exponentiating. To leading order in $D_\Delta$, the correction is given by $\frac{1}{2}\left(\langle S_{dis}^2\rangle-\langle S_{dis}\rangle^2\right)$ where the disconnected terms cancel.
After performing the integrals over the momentum-shell and transforming back to the real-space representation, we obtain
\begin{widetext}
\begin{equation}
\begin{aligned}
\frac{\langle\langle\langle S_{dis}^2\rangle_>\rangle\rangle-\langle\langle\langle S_{dis}\rangle_>^2\rangle\rangle}{2} =&-\frac{D_\Delta}{2} \int dx d\tau d\tau^\prime \xi_R^<\left(\tau\right)\xi_L^<\left(\tau\right)\xi_R^<\left(\tau^\prime\right)\xi_L^<\left(\tau^\prime\right)\\
&-\frac{2D_\Delta \Delta}{u_-^2\left(1+\left(\nicefrac{\Delta}{u_-\Lambda}\right)^2\right)} \left(\frac{1}{\Lambda^\prime}-\frac{1}{\Lambda}\right) \int dxd\tau i\xi_R^<\xi_L^< \\ & - \frac{D_\Delta}{2 u_-^2\left(1+\left(\nicefrac{\Delta}{u_-\Lambda}\right)^2\right)} \left(\frac{1}{\Lambda^\prime}-\frac{1}{\Lambda}\right) \int dxd\tau \left(\xi_R^<\partial_\tau \xi_R^< + \xi_L^< \partial_\tau \xi_L^<\right)\; .
\label{eq:RGContributionAssymDis}
\end{aligned}
\end{equation}
\end{widetext}

The last step in the RG procedure is to rescale the coordinates and fields. In momentum-space, we re-write $q\mapsto q e^{d\ell}$ for $q$ to restore the original cutoff $\Lambda$. The differentials $dx$, $d\tau$ correspondingly are multiplied by a factor $e^{d\ell}$ each, and $\xi$ are multiplied by $e^{y_\xi d\ell}$ where $y_\xi$ is their scaling dimension. In the clean model, $y_\xi = -\frac{1}{2}$; however, here there is a correction of order $D_\Delta$ required to compensate for the last term in Eq. (\ref{eq:RGContributionAssymDis}), adjusting the overall coefficient of the term $\xi\partial_\tau \xi$ in the effective action to have a coefficient unity.
Substituting these rescaling factors, the leading term with coupling to four Fermion fields becomes
\begin{widetext}
\begin{equation}
e^{(3+4y_\xi)d\ell} D_\Delta \int dx d\tau d\tau^\prime \xi_R\left(x,\tau\right)\xi_L\left(x,\tau\right)\xi_R\left(x,\tau^\prime\right)\xi_L\left(x,\tau^\prime\right)
\end{equation}
\end{widetext}
which gives the RG equation for $D_\Delta$. The equations for $\Delta$ and $u_-$ arise from the appropriate rescaling of the fields and coordinates in the last two terms of (\ref{eq:RGContributionAssymDis}). This concludes our derivation of Eq. (\ref{RG_eqn_as}) in the main text.

The above derivation relied on the ability to switch between real-space and momentum-space in a straightforward manner. This is useful for additional disorder terms that are quadratic in the free fields of $S_0$, such as $\delta U\left(x\right) \left(\partial_x \theta_+\right)^2$ in the symmetric sector. However, when there are non-quadratic operators involved, the procedure is more complicated as the coupling between fast and slow fields is tighter, and a simple representation of $\hat{O}$ in $\left(k,\omega\right)$-space is lacking.

To deal with this type of disorder terms, certain approximations will need to be implemented in the procedure of integrating the fast modes.
We employ the strategy described below for a general disorder term. Subsequently, we apply this approach to analyze the chemical potential disorder term $H_\eta$ [Eq. (\ref{eq:Hdis})].

We begin by splitting the double-time integral of Eq. (\ref{eq:BasicSdis2}) to two different terms, $\tau\approx\tau^\prime$ and $\tau\not\approx\tau^\prime$, where the former accounts for time-differences $\Delta\tau\equiv\tau-\tau^\prime$ within the short-time cutoff $\left(u\Lambda\right)^{-1}$:
\begin{equation}
\begin{aligned}
S_{dis}^2=&\int_{\tau\not\approx\tau^\prime} dx d\tau d\tau^\prime D_O \hat{O}\left(x,\tau\right)\hat{O}\left(x,\tau^\prime\right) \\
+&\int_{\tau\approx\tau^\prime} dx d\tau d\tau^\prime D_O \hat{O}\left(x,\tau\right)\hat{O}\left(x,\tau^\prime\right)\; .
\label{eq:SdisEqn}
\end{aligned}
\end{equation}
Generally, different local operators $\hat{O}_i$ are generated from local expansions of the term $\hat{O}\left(x,\tau^\prime\right)$:
\begin{equation}
\hat{O}\left(x,\tau^\prime\right)=\hat{O}\left(x,\tau\right)+\left(\Delta \tau\right)\partial_\tau \hat{O}\left(x,\tau\right)+\dots
\label{eq:local_expansion}
\end{equation}
One can expand to leading orders in $\Delta\tau$, resulting with a set of local operators (independent of $\tau^\prime$) multiplied by some function of $\Delta\tau$:
\begin{equation}
\begin{aligned}
S_{dis}^2=&\int_{\tau\not\approx\tau^\prime} dx d\tau d\tau^\prime D_O \hat{O}\left(x,\tau\right)\hat{O}\left(x,\tau^\prime\right) \\
+&\sum_i \int_{\tau\approx\tau^\prime} dx d\tau d\left(\Delta\tau\right) D_O F_i\left(\Delta\tau\right)\hat{O}_i\left(x,\tau\right)\; .
\end{aligned}
\end{equation}

We now turn to integrating over the fast modes. In the first term, we know that the correlation of the fast modes $\langle\hat{O}^>\left(\tau\right)\hat{O}^>\left(\tau^\prime\right)\rangle$ decays, so that under the approximation $\tau\not\approx\tau^\prime$ averaging over the fast modes will not depend on $\Delta\tau$. The second term is already composed only from local terms by construction. We therefore only need to calculate local expectation values. The result can be brought to the following structure:
\begin{equation}
\begin{aligned}
S_{dis}^{<2}=& e^{\gamma_O d\ell} \int_{\tau\not\approx\tau^\prime} dx d\tau d\tau^\prime D_O \hat{O}^<\left(x,\tau\right)\hat{O}^<\left(x,\tau^\prime\right) \\
+&\sum_i e^{\gamma_i d\ell} \int_{\tau\approx\tau^\prime} dx d\tau d\left(\Delta\tau\right) D_O F_i\left(\Delta\tau\right) \hat{O}_i^<\left(x,\tau\right)
\end{aligned}
\end{equation}
where $\gamma_i$ and $\gamma_O$ are related to the scaling dimensions of the operators, as will be seen in what follows.

To restore back the effective $S_{dis}^2$ in the slow modes sector to the form of (\ref{eq:RGStructure}), one must unite the $\tau\not\approx\tau^\prime$ and $\tau\approx\tau^\prime$ contributions to one term. A part of the local term is "absorbed" back in the non-local term, to reconstruct the structure of a disorder term. Following integration over $\Delta\tau$ we obtain

\begin{equation}
\begin{aligned}
S_{dis}^{<2}=&e^{\gamma_O d\ell} \int dx d\tau d\tau^\prime D_O \hat{O}^<\left(x,\tau\right)\hat{O}^<\left(x,\tau^\prime\right) \\
+&\sum_i \left(e^{\gamma_i d\ell}-e^{\gamma_O d\ell}\right) \int dx d\tau D_O C_i \hat{O}_i^<\left(x,\tau\right)
\end{aligned}
\label{eq:DisOslow}
\end{equation}
where $C_i\equiv\int d\left(\Delta\tau\right) F_i\left(\Delta\tau\right)$, in which the integral is bounded by the cutoff $\left(u\Lambda\right)^{-1}$ and yields a non-universal constant. The exact value of $C_i$ is not important -- only its sign and dependence on the parameters of the model. Note that $e^{\gamma_i d\ell}-e^{\gamma_O d\ell}\approx\left(\gamma_i-\gamma_O\right)d\ell$, so this can be understood as a correction to the coefficient of $\hat{O}_i$ (an operator that typically exists in the free action $S_0$) of order $d\ell$.

To complete the RG transformation, we have to restore the cutoff $\Lambda$. Similarly to the discussion of the quadratic case, $dx$ and $d\tau$ will each be multiplied by a factor of $e^{d\ell}$. In the limit $d\ell\to 0$, this rescaling can be neglected in the second line of Eq. (\ref{eq:DisOslow}). However, in the first, non-local term it yields an overall prefactor $e^{\left(3+\gamma_O\right)}$;
we interpret the resulting coefficient as the renormalized disorder. Noting that $\gamma_O<0$, the exponent $y_{_{D_O}}\equiv 3+\gamma_O$ is the scaling dimension of the disorder operator, which will determine the condition for it to be relevant. The second, local term provides a set of corrections to the parameters of $S_0$.

The last step is re-exponentiation -- once again leading to subtraction of the disconnected term $\langle S_{dis}\rangle^2$. This yields the final form Eq. (\ref{eq:RGStructure}).

To demonstrate the general procedure described above, we now briefly review the analysis the disorder term $H_g$ [Eq. (\ref{eq:Hdis_g})]. The operator in this case is $\hat{O}=\cos\sqrt{8}\theta_+$. We will use intermediate calculations that match appendix E of Ref. \onlinecite{27 Giamarchi 04}, and the final result will be identical to Ref. \onlinecite{GS}.
Averaging over the fast modes we have
\begin{equation}
\langle\hat{O}\left(x,\tau\right)\rangle_> = e^{-\frac{2}{K}} \hat{O}^<\left(x,\tau\right)
\end{equation}
and therefore $\gamma_O=-\frac{4}{K}$, so the scaling of $D_g$ is $\frac{d D_g}{d\ell}=\left(3-\frac{4}{K}\right)D_g$.

Looking at the short-range regime $\tau\approx\tau^\prime$, the product $\hat{O}\left(\tau\right)\hat{O}\left(\tau^\prime\right)$ can be simplified using trigonometrical identities and the expansion Eq. (\ref{eq:local_expansion}). The result yields two local terms in the leading orders:
\begin{equation}
\begin{aligned}
\hat{O}_1\left(\tau\right)=&\cos\sqrt{32}\theta_+\\
\hat{O}_2\left(\tau\right)=&\left(\partial_\tau \theta_+\right)^2\; .
\end{aligned}
\end{equation}
The operator $\hat{O}_1$ is not very interesting, as its dimension is very low, $2-\frac{8}{K}$ and so it is irrelevant in our regime of interest. The operator $\hat{O}_2$, on the other hand, will lead to the corrections to $u_{_+}$ and $K_{_+}$ as they appear in (\ref{RG_BosonI})-(\ref{RG_Fermion})

We next turn to apply this approach for the analysis of $H_\eta$ [Eq. (\ref{eq:Hdis})], where $\hat{O}=\cos\left(\sqrt{2}\theta_+\right)\cos\left(\sqrt{2}\theta_-\right)$. As we follow the same procedure, to linear order in $D_\eta$ one straightforwardly obtains the corrections to various terms which couple to local operators ${\hat{O}_i}$. However, the RG transformation of the disorder term itself poses a challenge: as long as the gap $\Delta$ in the anti-symmetric sector is finite, the correlations of $\cos\left(\sqrt{2}\theta_-\right)$ never decay as a power-law for $\Delta\tau\rightarrow\infty$. Rather, employing the decomposition
\begin{equation}
\cos\sqrt{2}\theta_-=\langle\cos\sqrt{2}\theta_-\rangle + :\cos\sqrt{2}\theta_-:
\label{eq:OP_fluc}
\end{equation}
the second term has exponentially decaying correlations. For $V>0$, the first term vanishes ($\cos\sqrt{2}\theta_-$ couples to the disorder field in the Ising representation \cite{27 Gogolin Nersesyan 98}); hence $\hat{O}$ is exponentially irrelevant to the present order in the perturbative expansion of $H_\eta$.

To derive RG flow equations for $D_\eta$, we therefore need to consider the next order in the perturbative expansion. This yields a disorder term coupling to the operator $\cos\sqrt{8}\theta_-$, which has a simple representation in terms of either the Fermion fields $\xi_{R,L}$, or the massive Bosonic field $\theta_-$. Indeed, the fourth order term $S_{dis}^4$ (with a coefficient proportional to $D_\eta^2$) contains
a specific {\it four-point} combination $\hat{O}\left(x,\tau\right)\hat{O}\left(x,\tau\right)\hat{O}\left(x,\tau^\prime\right)\hat{O}\left(x,\tau^\prime\right)$ which possesses
power-law decaying correlations. Using $\cos^2\left(\sqrt{2}\theta_-\right)=\frac{1}{2}[1+\cos\left(\sqrt{8}\theta_-\right)]$, this contributes several terms: some of them are local and can be interpreted as corrections to $\delta$ and $g$ (the latter renormalizing $D_g$ as well);
the leading non-local (``disorder-like") term is therefore associated with the operator $\hat{\tilde{O}}=\cos\left(\sqrt{8}\theta_+\left(x,\tau\right)\right)\cos\left(\sqrt{8}\theta_-\left(x,\tau\right)\right)$, with a coefficient $\propto D_\eta^2$.

Proceeding with the analysis of the latter disorder term is made possible by implementing the approximate representation of $\cos\left(\sqrt{8}\theta_\pm\right)$ in terms of the free fields, and explicitly evaluating $\langle\;\rangle_>$. In particular, the operator $\cos\left(\sqrt{8}\theta_+\right)$ is already included in $H_g$ and yields the same scaling exponent; the scaling dimension of $\cos\left(\sqrt{8}\theta_-\right)$ can be inferred from either the Fermionic or the massive Bosonic representations, depending on the value of $K_{_-}$ (see main text). This yields the following scaling dimensions:
\begin{equation}
y_{_{D_\eta}}=
\begin{cases}
\frac{1}{2}-\frac{2}{K_{_+}}&\text{Fermionic behavior}\\
\frac{3}{2}-\frac{2}{K_{_+}}-\frac{2}{K_{_-}\left(1+\delta^2\right)}&\text{Bosonic behavior}
\end{cases}
\end{equation}
with $\delta$ the dimensionless gap in the Bosonic regime. Note that in any case, $D_\eta$ is always less relevant than $D_g$; hence the emergence of a disordered insulating phase is dominated by the critical value of $D_g$, and is only indirectly dependent on $D_\eta$ via the corrections it generates to the other parameters.

We now comment on the contribution to the RG equations coming from linear order in $D_\eta$. These arise from corrections to the coefficients of the following local operators $\left\{\hat{O}_i\right\}$:
\begin{equation}
\begin{aligned}
\hat{O}_1\left(\tau\right)=&\left(\partial_\tau \theta_+\right)^2 \\
\hat{O}_2\left(\tau\right)=&\cos\sqrt{8}\theta_- \\
\hat{O}_3\left(\tau\right)=&\left(\partial_\tau \theta_-\right)^2\; ;
\end{aligned}
\end{equation}
these result in the contribution of $\mathcal{D}_\eta$ to Eqs. (\ref{RG_BosonI}) and (\ref{RG_Fermion}).

We finally note that in the regime where the anti-symmetric sector is a gapped superconductor, any operator which contains non-trivial factors of $\cos\gamma\theta_-$ (with arbitrary $\gamma$) is exponentially irrelevant, and contributes nothing to any order in $D_\eta$. The only contributions come from terms in the expansion that couple only to $\theta_+$, and therefore, at least to leading (second) order, the effect of $D_\eta$ is just creating a shift in $D_g$:
\begin{equation}
D_g\to D_g^{eff} = D_g+\alpha D^2_\eta\; .
\end{equation}
For this reason, deep in the SC phase $H_\eta$ can be ignored altogether and the $\pm$-sectors are effectively decoupled.

\section{Analytic Solutions of the RG Equations}
\label{app:analytical}

The set of equations described in Eq. (\ref{RG_Fermion}) is coupled, and an analytic solution will be complicated if it even exists. However, some special cases can be helpful to understand the type of flow expected in each phase.

\subsection{Quadratic Disorder}
\label{D_UsAnalysis}

As discussed in Sec. \ref{sec:RGanalysis}A of the main text, the disorder in quadratic terms of the symmetric (gapless) sector $\delta U\left(x\right) \left(\partial_x\theta_+\right)^2$, $\delta \rho_s\left(x\right) \left(\partial_x\phi_+\right)^2$ is always irrelevant, and therefore we have not discussed it in detail in the main text. However, in the presence of this disorder alone one can exactly solve the equations, and understand its effect on the system.

The equations for the disorder itself are simple:
\begin{equation}
\frac{d\mathcal{D}_{U/s}}{d\ell}=-\mathcal{D}_{U/s}
\end{equation}
so the solution is just an exponent
\begin{equation}
\mathcal{D}_{U/s}\left(\ell\right)=\mathcal{D}_{U/s}^0 e^{-\ell}
\end{equation}
This can be substituted in the equations for $K_{_+},u_{_+}$:
\begin{equation}
\begin{aligned}
\frac{dK_{_+}}{d\ell}=&4\left(\mathcal{D}_s^0K_{_+}^2-\frac{\mathcal{D}_U^0}{K_{_+}^2}\right)e^{-\ell} K_{_+}\\
\frac{du_{_+}}{d\ell}=&-4\left(\mathcal{D}_s^0K_{_+}^2+\frac{\mathcal{D}_U^0}{K_{_+}^2}\right)e^{-\ell}u_{_+}
\end{aligned}
\end{equation}

Generally solving these equations is hard, but to leading order in $\mathcal{D}_{U/s}$ the renormalization of $K_{_+}$ inside the round brackets can be neglected. The resulting flow is of the form $\frac{dQ}{d\ell}=A e^{-\ell}Q$, with the solution $Q(\ell)=Q_0 e^{A\cdot\left(1-e^{-\ell}\right)}$. The asymptotic values (at $\ell\rightarrow\infty$) are
\begin{equation}
\begin{aligned}
u_{_+}\left(\infty\right)=&u_{_+} e^{-4\left(\mathcal{D}_s^0K_{_+}^2+\frac{\mathcal{D}_U^0}{K_{_+}^2}\right)} \\
K_{_+}\left(\infty\right)=&K_{_+} e^{4\left(\mathcal{D}_s^0K_{_+}^2-\frac{\mathcal{D}_U^0}{K_{_+}^2}\right)}\; .
\end{aligned}
\end{equation}
As this flow converges very fast, we generally ignored it and assumed the values of $K_{_+}$ and $u_{_+}$ we were using are the stable ones (in terms of $\mathcal{D}_{U/s}$).

One should also consider the effect of these disorder terms on the antisymmetric sector. Generally the effect should be similar, but adding a factor of $\left(1+\delta^2\right)$ complicates the calculations. We can note that qualitatively, the effect on the anti-symmetric sector will be weaker. This is what gives us the option to tune the ratio $\frac{K_{_+}}{K_{_-}}$ to be above or below $1$, independently of the sign of $V$ (which does constrain the bare values) -- a strong $\mathcal{D}_s$ term will push the ratio up, and a strong $\mathcal{D}_U$ term will push it down.

\subsection{Antisymmetric Sector}
\label{app:DminusAnalysis}

The disorder in $\Delta$ in the antisymmetric sector is special, as $\mathcal{D}_-$ always diverges: $\frac{d\mathcal{D}_-}{d\ell}>\mathcal{D}_-$, which indicates an exponential divergence. Yet, there is an ordered phase in the sector:
as $\mathcal{D}_-$ is the variance in the dimensionless gap $\delta$, if $\sqrt{\mathcal{D}_-}\ll \delta$ the randomness in the gap is actually small compared to the average gap $\delta$, and the clean limit is effectively recovered. Indeed, in the RG equations $\mathcal{D}_-$ is always divided by $1+\delta^2$.

As diverging $\mathcal{D}_-$ is not a good indicator for the nature of the phase, one should look instead on the "normalized disorder" parameter $\mathcal{\tilde{D}_-}=\frac{\mathcal{D}_-}{1+\delta^2}$ defined in Sec. \ref{sec:RGanalysis}B, which obeys Eq. (\ref{RG_eqn_as_normal}) in the case $\mathcal{D}_\eta=0$. The RG equations for the velocity $u_{_-}$ and $\delta$ become
\begin{equation}
\frac{du_{_-}}{d\ell}=-\mathcal{\tilde{D}_-}u_{_-}\; ,\quad\frac{d\delta}{d\ell}=(1-2\mathcal{\tilde{D}_-})\delta\; .
\end{equation}
Therefore, diverging $\mathcal{\tilde{D}_-}$ will lead to $u_{_-},\delta\to 0$, which characterizes a disordered phase.
However, from Eq. (\ref{RG_eqn_as_normal}) it is clear that $\mathcal{\tilde{D}_-}=\frac{3}{16}$ is a turning point. If $\mathcal{\tilde{D}_-}>\frac{3}{16}$, $\frac{d\mathcal{\tilde{D}_-}}{d\ell}>0$ and the disorder will diverge. On the other hand, if $\mathcal{\tilde{D}_-}<\frac{3}{16}$, there is a critical value of $\delta$ above which $\mathcal{\tilde{D}_-}$ will flow downwards and turn irrelevant: $\delta_c=\frac{1}{1-\nicefrac{16}{3}\mathcal{\tilde{D}_-}}$.

As a rule of thumb, for low initial values there is no strong coupling between $\delta$ and $\mathcal{\tilde{D}_-}$ (no linear contributions from one parameter on the other) and the main question is what will happen first -- either $\delta$ will reach a value of order $1$, or $\mathcal{\tilde{D}_-}$ will approach the critical value $\frac{3}{16}$ or close to it. As $\delta$ and $\mathcal{\tilde{D}_-}$ scale the same close to the point $\left(\delta=0,\mathcal{\tilde{D}_-}=0\right)$, one expects linear critical lines yielding the V-shape $\mathcal{\tilde{D}_-}\propto\left|\delta\right|$, as can be seen in Fig. \ref{fig:phase_diagram_AS}. We finally note that to get the right panel of this figure, one may use the explicit $K_{_-}$-dependence of $\Delta$ [e.g. Eq. (\ref{eq:mass2K_-}), or Eqns. (25)-(27) in \ref{Mazo2014}]; this leads to the curving of the phase-boundary in the top and bottom of the figure.

\section{Bosons with a random mass}
\label{app:OtherDis}

In this Appendix we consider a type of disorder which was not discussed in the main text: randomness in the gap characterizing the massive Bosonic regimes of the antisymmetric sector. Specifically, deep in the superconducting phase, one can write the following effective Hamiltonian to describe fluctuations in the relative phase $\phi_-$:
\begin{equation}
H_-=\frac{u_{_-}}{2\pi} \int dx \left[K_{_-}\left(\partial_x\theta_-\right)+\frac{1}{K_{_-}}\left(\partial_x\phi_-\right)^2 + \Delta^2 \phi_-^2 \right]
\end{equation}
where we have replaced $g_\phi \cos\phi_-$ with the gap term $\Delta^2\phi_-^2$ following the gap equation for a sine-Gordon model \cite{27 Giamarchi 04}:
\begin{equation}
\frac{\Delta}{u_{_-}\Lambda}=\left(\frac{2\pi^2K_{_-}g_\phi}{\Lambda^2u_{_-}}\right)^{\frac{1}{2-\nicefrac{K_{_-}}{2}}}\; .
\label{eq:mass2K_-}
\end{equation}
As $\Delta$ is affected by various parameters like $u_{_-}$, $K_{_-}$ and $g_\phi$, once either of them develops randomness it must also fluctuate in space. We therefore replace $\Delta\mapsto\Delta+\delta\Delta(x)$, with $\left<\left<\delta\Delta(x)\delta\Delta(x^\prime)\right>\right>=D_\phi \delta\left(x-x^\prime\right)$.

Defining the dimensionless parameters $\mathcal{D}_\phi=\frac{\pi D_\phi}{u_{_-}^2\Lambda}$, $\delta=\frac{\Delta}{u_{_-}\Lambda}$ and performing an analysis along the lines described in App. \ref{app:RG}, the RG equations to leading order in $\mathcal{D}_\phi$ are given by
\begin{equation}
\begin{aligned}
\frac{d\delta}{d\ell}=&\delta\left(1-\frac{8K_{_-}\mathcal{D}_\phi}{1+\delta^2}\right) \\
\frac{d\mathcal{D}_\phi}{d\ell}=&\mathcal{D}_\phi
\end{aligned}
\end{equation}
which can flow either to $\left(\delta=0,\mathcal{D}_\phi=\infty\right)$ or to $\left(\delta=\infty,\mathcal{D}_\phi=\infty\right)$. The former is a disordered superconductor, with strong randomness in the gap -- some kind of vortex-glass, perhaps \cite{Fisher 89}; the exact nature can not be determined from this approximate, perturbative approach. The latter case obeys $\frac{\mathcal{D}_\phi}{\delta}\to Const.<1$, which means that the width of the distribution of $\delta$ gets smaller compared to $\delta$ itself.

This behavior exists, of course, in the Bosonic insulator as well. However, it will never change the \textit{structure} of the phase diagram, as it does not affect the Fermionic intermediate sector. The superconducting phase, the intermediate disordered phase and the intermediate ordered phase will all exist, and the main effect of this disorder in $\theta$ will be just inside the disordered insulating phase, where $\mathcal{D}_\eta$ is dominating anyway and drives the system to a disordered insulator state.


\vspace{5cm}


\begin{thebibliography}{10}


\bibitem{1 Hebard 93} For a review and extensive references, A. F.
Hebard, in Strongly Correlated Electronic Materials (The Los Alamos
Symposium 1993), edited by K. S. Bedell, Z. Wang, D. E. Meltzer\c{ }A.
V. Balatsky, and E. Abrahams, Addison Wesley (1994), p. 251.

\bibitem{1 Sondhi Girvin 97}S. L. Sondhi, S. M. Girvin, J. P. Carini
and D. Shahar,
Rev. Mod. Phys. \textbf{69}, 315 (1997).

\bibitem{1 Goldman Markovic 98}A. M. Goldman and N. Markovic, Physics
Today \textbf{51}, 39 (1998).

\bibitem{3 Sachdev 19}S. Sachdev, \emph{Quantum Phase Transitions}
(Cambridge University Press (1999)).


\bibitem{18 Fisher-Weichman 89} M. P. A. Fisher, P. B. Weichman,
G. Grinstein and D. S. Fisher, Phys. Rev. B \textbf{40}, 546 (1989).

\bibitem{18 Vojta-Crewse 16} T. Vojta, J. Crewse, M. Puschmann, D.
Arovas and Y. Kiselev, Phys. Rev. B \textbf{94}, 134501 (2016).

\bibitem{PRLB Jul18 32 Fisher 90}M. P. A. Fisher and D. H. Lee, Phys.
Rev. B \textbf{39}, 2756 (1989), M. P. A. Fisher, Phys. Rev. Lett. \textbf{65}, 923 (1990).

\bibitem{PRLB Jul18 37 Fisher-Grinstein 90} M. P. Fisher, G. Grinstein,
and S. M. Girvin, Phys. Rev. Lett. \textbf{64}, 587 (1990).

\bibitem{PRLB Jul18 38 Sorensen-Wallin 92} E. S. S\o rensen, M. Wallin,
S. M. Girvin, and A. P. Young, Phys. Rev. Lett. \textbf{69}, 828 (1992).

\bibitem{PRLB Jul18 39 Cha-Girvin 39} M. C. Cha and S. M. Girvin, Phys. Rev. B \textbf{49}, 9794 (1994).

\bibitem{PRLB Jul18 40 Prokof'ev-Svistunov} N. Prokof\textquoteright ev
and B. Svistunov, Phys. Rev. Lett. \textbf{92}, 015703 (2004).

\bibitem{Altman Kafri Polkovnikov Refael 10}E. Altman, Y. Kafri,
A. Polkovnikov and G. Refael, Phys. Rev. Lett. \textbf{93}, 150402
(2004); E. Altman, Y. Kafri, A. Polkovnikov and G. Refael,
Phys. Rev. Lett. \textbf{100}, 170402 (2008); E. Altman, Y. Kafri,
A. Polkovnikov and G. Refael, Phys. Rev. B \textbf{81}, 174528 (2010).

\bibitem{19 Phillips 03} P. Phillips, Science \textbf{302}, 243 (2003).

\bibitem{19 Kapitulnik-Kivelson arXiv} A. Kapitulnik, S. A. Kivelson
and B. Spivak, Rev. Mod. Phys. \textbf{91}, 011002 (2019); and refs. therein.

\bibitem{20 Mulligan-Raghu 16} M. Mulligan and S. Raghu, Phys. Rev. B \textbf{93}, 205116 (2016).

\bibitem{20 Goldman-Mulligan 17} H. Goldman, M. Mulligan, S. Raghu,
G. Torroba and M. Zimet, Phys. Rev. B \textbf{96}, 245140 (2017).


\bibitem{22 Orignac-Giamarchi 98} E. Orignac and T. Giamarchi, Phys. Rev. B \textbf{57}, 11713 (1998).

\bibitem{22 Dhar-Maji 12} A. Dhar, M. Maji, T. Mishra, R. V. Pai,
S. Mukerjee and A. Paramekanti, Phys. Rev. A \textbf{85}, 041602(R)
(2012).

\bibitem{22 Tokuna-Georges 14} A. Tokuno and A. Georges, New J. of Phys. \textbf{16}, 073005 (2014).

\bibitem{22 Ristivojevic-Petkovic 14} Z. Ristivojevic, A. Petkovic,
P. Le Doussal and T. Giamarchi, Phys. Rev. B \textbf{90}, 125144 (2014).

\bibitem{28 Atzmon Shimshoni 11}Y. Atzmon and E. Shimshoni, Phys. Rev. B \textbf{83}, 220518(R) (2011).

\bibitem{Mazo2014}
V. Mazo, C.-W. Huang, E. Shimshoni, S. T. Carr, and H. A. Fertig, Phys. Rev. B {\bf 89}, 121411(R) (2014); V. Mazo, C.-W. Huang, E. Shimshoni, S. T. Carr, and H. A. Fertig, Phys. Scr. T {\bf 165}, 014019 (2015).

\bibitem{randomIsing}
R. Shankar and G. Murthy, Phys. Rev. B {\bf 36}, 536 (1987); D. S. Fisher, Phys. Rev. Lett. {\bf 69}, 534 (1992); D. S. Fisher, Phys. Rev. B {\bf 51}, 6411 (1995).

\bibitem{KosterlitzThouless}
J. M. Kosterlitz and D. J. Thouless, J. Phys. C: Solid State {\bf 6}, 1181 (1973)

\bibitem{Berezinskii}
V. L.  Berezinskii, Sov. Phys. JETP {\bf 34}, 610 (1972).

\bibitem{Nandini_etal}
A. Ghosal, M. Randeria and N. Trivedi, Phys. Rev. B {\bf 65}, 014501 (2001); A. Datta, A. Banerjee, N. Trivedi and A. Ghosal, arXiv:2101.00220.

\bibitem{Sitte}
M. Sitte, A. Rosch, J. S. Meyer, K. A. Matveev and M. Garst, Phys. Rev. Lett. {\bf 102}, 176404 (2009).

\bibitem{29 Huijse Bauer 15}L. Huijse, B. Bauer and E. Berg, Phys. Rev. Lett. \textbf{114}, 090404 (2015).

\bibitem{29 Alberton Ruhman 17}O. Alberton, J. Ruhman, E. Berg and
E. Altman, Phys. Rev. B \textbf{95}, 075132 (2017).

\bibitem{27 Giamarchi 04}
T. Giamarchi, \emph{Quantum Physics in One Dimension} (Oxford University Press, 2004).

\bibitem{SDSG}
P. Lecheminant, A. O. Gogolin and A. A. Nersesyan, Nucl. Phys. B {\bf 639}, 502 (2002).

\bibitem{27 Gogolin Nersesyan 98} A. O. Gogolin, A. A. Nersesyan and A.
M. Tsvelik, \emph{Bosonization and Strongly Correlated Systems} (Cambridge
University Press, 1998).

\bibitem{GS}
T. Giamarchi and H. J. Schulz, Phys. Rev. B {\bf 37}, 325 (1988).

\bibitem{Fisher 89}
M.P.A. Fisher, Phys. Rev. Lett. {\bf 62}, 1415 (1989)

\end{thebibliography}
\end{document}